\documentclass[9pt, letter, conference]{IEEEtran}
\IEEEoverridecommandlockouts
\usepackage{hyperref}
\usepackage{amsmath,amsfonts}
\usepackage{colortbl}
\usepackage{array} 
\usepackage{multirow}
\usepackage{multicol}
\usepackage{amsthm}
\usepackage{listings}
\usepackage{amsmath}
\usepackage[linesnumbered,ruled,vlined]{algorithm2e}
\usepackage{subcaption}  
\usepackage{array,tabularx}
\newcolumntype{C}[1]{>{\centering\arraybackslash}p{#1}}
\usepackage{booktabs}       
\renewcommand{\arraystretch}{1.2} 

\usepackage{graphicx}
\usepackage{pgfplots}
\usepackage{textcomp}
\usepackage{xcolor}
\usepackage{cleveref}
\usepackage{makecell}
\usepackage{booktabs}
\usepackage{comment}
\usepackage{tabularx}
\usepackage{multirow}
\usepackage{bigdelim}
\usepackage{bm}
\usepackage{url}
\usepackage{xspace}
\usepackage{pifont}
\usepackage{arydshln}
\usepackage{verbatim}
\usepackage[referable]{threeparttablex}
\usepackage{calc} 
\usepackage{extarrows}
\usepackage{color}
\usepackage{float}
\usepackage{subcaption}
\usepackage{nicematrix}
\usepackage{geometry}
\usepackage{balance}
\usepackage[utf8]{inputenc}
\usepackage{hhline}
\usepackage{booktabs} 
\usepackage{tabularx} 
\usepackage{tikz}
\usetikzlibrary{shapes.geometric, arrows}
\tikzstyle{startstop} = [rectangle, rounded corners, minimum width=3cm, minimum height=1cm,text centered, draw=black]
\tikzstyle{io} = [trapezium, trapezium left angle=70, trapezium right angle=110, minimum width=3cm, minimum height=1cm, text centered, draw=black]
\tikzstyle{process} = [rectangle, minimum width=3cm, minimum height=1cm, text centered, draw=black]
\tikzstyle{arrow} = [thick,->,>=stealth]

\usepackage{filecontents}                                  
\usepackage{pgfplots}
\usepackage{pgfplotstable}
\usepackage{scalefnt}

\definecolor{USTgold}{RGB}{153,102,0}
\definecolor{USTyellow}{RGB}{204,153,0}
\definecolor{USTyellowlight}{RGB}{255,212,0}
\definecolor{USTorange}{RGB}{255,166,26}
\definecolor{USTpink}{RGB}{255,157,157}
\definecolor{USTblue}{RGB}{0,51,102}
\definecolor{USTmiddle}{RGB}{0,116,188}
\definecolor{USTlight}{RGB}{99,202,225}
\definecolor{USTgray}{RGB}{204,204,204}
\definecolor{USTred}{RGB}{237,27,47}
\definecolor{USTdarkred}{RGB}{124,35,72}

\definecolor{CUHKorange}{RGB}{244,106,18} 
\definecolor{CUHKblue}{RGB}{0,111,190}    
\definecolor{CUHKgreen}{RGB}{0,127,128}   
\definecolor{CUHKred}{RGB}{228,46,36}     
\definecolor{CUHKyellow}{RGB}{198,148,34} 
\definecolor{CUHKdark}{RGB}{114,44,114}   
\definecolor{CUHKmiddle}{RGB}{144,44,144} 
\definecolor{CUHKlight}{RGB}{167,44,167} 
\definecolor{lightblue}{RGB}{223, 235, 247}

\newcommand{\deftitle}[0]{{\texttt{CRISTAL}}\xspace}

\iftrue
\def\BibTeX{{\rm B\kern-.05em{\sc i\kern-.025em b}\kern-.08em
    T\kern-.1667em\lower.7ex\hbox{E}\kern-.125emX}}

\fi

\iftrue
\fi

\iftrue

\usepackage{titlesec}
\titlespacing\section{2pt}{5pt plus 1pt minus 1pt}{0pt plus 1pt minus 1pt}
\titlespacing\subsection{2pt}{5pt plus 1pt minus 1pt}{0pt plus 1pt minus 1pt}
\titlespacing\subsubsection{2pt}{5pt plus 1pt minus 1pt}{2pt plus 1pt minus 1pt}
\usepackage[inline]{enumitem}
\setlist{leftmargin=5.08mm}
\fi

\iftrue
\fi

  \newcolumntype{B}{>{\bfseries\centering\arraybackslash\hsize=0.16\hsize}X}
  \newcolumntype{Y}{>{\centering\arraybackslash\hsize=0.84\hsize}X}
\usepackage{tabularx}
\usepackage{ragged2e}  
\newcolumntype{Y}{>{\RaggedRight\arraybackslash}X}
\usepackage{cite}

\usepackage{cleveref}
\usepackage{xspace}

\newcounter{question}
\newcommand{\question}[1]{%
    \par\vspace{0.5em}
    \refstepcounter{question}
    \noindent\textbf{\underline{Q\arabic{question}:}}~\textit{#1}
    \par\vspace{0.5em}
}

\def\BibTeX{{\rm B\kern-.05em{\sc i\kern-.025em b}\kern-.08em
    T\kern-.1667em\lower.7ex\hbox{E}\kern-.125emX}}
\newlist{myitemize}{itemize}{1}
\setlist[myitemize]{
  label=\textbullet, 
}

\usepackage{etoolbox}
\makeatletter
\patchcmd{\@makecaption}
  {\scshape}
  {}
  {}
  {}
\makeatother
\DeclareMathOperator*{\argmax}{arg\,max}
\def\BibTeX{{\rm B\kern-.05em{\sc i\kern-.025em b}\kern-.08em
    T\kern-.1667em\lower.7ex\hbox{E}\kern-.125emX}}

\begin{document}


\title{\LARGE Revisit Choice Network for Synthesis and Technology Mapping}





%


\author{\IEEEauthorblockN{ Chen Chen}
\IEEEauthorblockA{\textit{University of Maryland} \\
College Park, MD, US \\
cchen099@umd.edu}
\and
\IEEEauthorblockN{Jiaqi Yin}
\IEEEauthorblockA{\textit{University of Maryland} \\
College Park, MD, US \\
jyin629@umd.edu}
\and
\IEEEauthorblockN{Cunxi Yu}
\IEEEauthorblockA{\textit{University of Maryland} \\
College Park, MD, US \\
cunxiyu@umd.edu}

}

\maketitle

\begin{abstract}

Choice network construction is a critical technique for alleviating structural bias issues in Boolean optimization, equivalence checking, and technology mapping. Previous works on lossless synthesis utilize independent optimization to generate multiple snapshots, and use simulation and SAT solvers to identify functionally equivalent nodes. These nodes are then merged into a subject graph with choice nodes. However, such methods often neglect the quality of these choices—raising the question of whether they truly contribute to effective technology mapping.
This paper introduces \deftitle, a novel methodology and framework to constructing Boolean choice networks. Specifically, \deftitle introduces a novel flow of choice network-based synthesis and mapping, includes representative logic cone search, structural mutation for generating diverse choice structures via equality saturation, and priority-ranking choice selection along with choice network construction and validation. Through these techniques, \deftitle constructs fewer but higher-quality choices.
Our experimental results demonstrate that \deftitle outperforms the state-of-the-art Boolean choice network construction implemented in ABC in the post-mapping stage, achieving average reductions of \textbf{3.85\%/8.35\%} (area/delay) in delay-oriented mode, \textbf{0.11\%/2.74\%} in area-oriented mode, and a \textbf{63.77\%} runtime reduction on large-scale cases, across a diverse set of combinational circuits from the IWLS 2005, ISCAS'89, and EPFL benchmark suites.

\end{abstract}


\section{Introduction}
\par 
The concept of choice network was pioneered to address optimization limitations in Electronic Design Automation (EDA). The concept of choice network-embedding multiple alternative implementations (choice nodes) into a unified And-Inverter Graph (AIG)~\cite{aig2006} representation, enabling technology mapping algorithms to explore diverse structural realizations of the same logic~\cite{chatterjee2006reducing,chatterjee2007algorithms}, with wide range of applications in logic optimization, technology mapping, and physical-aware optimizations. Unlike traditional flows that map a single fixed topology, choice networks preserve optimization history by integrating equivalent variants generated through independent synthesis runs~\cite{mishchenko2006improvements}. This paradigm shift allows downstream tools to dynamically select the most library-friendly structure during mapping, thereby mitigating structural bias~\cite{chatterjee2006reducing}.


Despite their foundational role in enabling structural optimization, traditional choice node generation in EDA often relies on heuristic, locality-driven methods that miss global optimization opportunities. While some state-of-the-art approaches, such as Boolean matching~\cite{booleanmatching}, help reduce local structural bias by directly comparing logic functions, more advanced methods like lossless synthesis~\cite{mishchenko2006improvements, chatterjee2006reducing} attempt to address global structural bias by preserving multiple independently optimized representations and identifying functionally equivalent nodes using random simulation and SAT-based equivalence checking~\cite{sat}. It then strategically injects AIG subgraphs corresponding to the verified equivalent nodes as choice nodes into a unified AIG with choice nodes representation. A choice structure represents an alternative subgraph that is functionally equivalent but structurally distinct from its counterpart in the main graph. However, this approach has a significant limitation: the generated choice nodes rely solely on functional equivalent nodes produced by different independent optimization flows.
Based on our detailed analysis of the choice structures produced by the state-of-the-art implementation in \texttt{ABC}, we observe that these choices typically represent only minimal variations of the main graph's subgraphs. They are largely the result of local rewriting, with only small differences in a few nodes, while the majority of the structure remains identical. This inherently limits their ability to effectively address the problem of global structural bias.
Furthermore, we observe that simply increasing the number of choices is not inherently beneficial. This raises an important question: Can we rethink the process of generating choices? Instead of passively relying on the detection of functionally equivalent nodes during independent optimization --  could we instead actively and purposefully construct choice structures for the critical subgraphs that have a greater impact on improving mapping quality?

{Equality saturation, an innovative optimization technique, employs a non-destructive rewriting strategy through e-graphs to enable comprehensive Pareto-optimal design space exploration. This approach has seen broad adoption in Electronic Design Automation (EDA), proving particularly valuable for complex challenges in logic synthesis \cite{ustun2022impress,chen2024syn,coward2022automatic,chen2025emorphicscalableequalitysaturation} and formal verification \cite{coward2023datapath,yin2025boole}~\cite{coward2023automating}. Beyond its role in optimization, equality saturation inherently supports structural exploration by preserving diverse suboptimal implementations. These implementations exhibit significant divergence from the original logic cones, forming structural variants with the potential to mitigate global structural bias in technology mapping. Despite its natural alignment with the concept of choice networks—where multiple equivalent structural or functional options are maintained—equality saturation has not yet been leveraged for explicit choice network construction in EDA.}

To address this bottleneck, we have designed an efficient choice network construction framework called \textbf{CRISTAL}\footnote{It is open-sourced at \href{https://github.com/chestercc1997/Cristal}{https://github.com/chestercc1997/Cristal}}. This framework incorporates the following key components:
\begin{itemize}
    \item This work presents a fresh perspective on choice network construction by leveraging a formal representation of Boolean networks through Boolean-constrained \textit{equality saturation} and \textit{e-graphs} \cite{egg} to generate structural variants, complementing conventional simulation and SAT-solving techniques.

    \item This work realizes this novel idea by an end-to-end choice network framework, \textbf{CRISTAL}, featuring three key techniques: \textbf{representative cone pruning}, \textbf{hybrid structural mutation}, and \textbf{lightweight choice cones filtering}. Consequently, \deftitle achieves more effective exploration with fewer but higher-quality choices, alleviating the shortcomings of traditional quantity-driven methods.
    \item \deftitle is evaluated with comprehensive experimental setups, comparing to state-of-the-art choice network construction techniques implemented in \texttt{ABC}~\cite{aig2006}. On IWLS2005, EPFL~\cite{amaru2015epfl}, and ISCAS89~\cite{Brglez1989} benchmarks, our results demonstrate \textbf{3.85\%/8.35\%} (area/delay) reductions in delay-oriented mode and \textbf{0.11\%/2.74\%} reductions in area-oriented mode, with a \textbf{63.77\%} runtime reduction on large-scale cases in the post-mapping stage compared to conventional approaches.
    
    
\end{itemize}

\section{Preliminaries}
\begin{table*}[!t]
  \caption{Terminology Definitions for AIG with Choices}
  \label{terminology}
  \centering
  \renewcommand{\arraystretch}{1.4}
  \setlength{\tabcolsep}{6pt} 
  \small
  \begin{tabular}{@{}>{\bfseries\centering}p{4cm} >{\centering\arraybackslash}p{\dimexpr0.98\linewidth-4cm-4\tabcolsep\relax}@{}}
    \toprule
    Term & Definition \\
    \midrule

    Equivalence Class
    & A set of functionally equivalent nodes $\{n_1, n_2, \dots\}$ such that $f(n_i) = f(n_j)$ for all $i,j$. \\

    Representative Node
    & The primary node in an equivalence class, taken from the subject graph.  \\

    Choice Node
    & Any other node in the equivalence class besides the representative. \\

    Transitive Fanin (\textit{TFI})
    & All nodes that can be reached via backward traversal of fanin edges from a given node. Denoted $\mathcal{T}(r)$. \\

    Maximum Fanout-Free Cone (\textit{MFFC})~\cite{cong}
    & The largest subgraph within $\mathcal{TFI}(r)$ where every internal node $m \neq r$ has strictly internal fanout connections ($fanout(m) \subseteq \mathcal{M}_{\text{mffc}}(r)$) and all paths from the subgraph to primary outputs must route through $r$. \\

    Choice Node 
    & A node $c$ in an equivalence class where $c \neq r$ (representative node) and $\mathcal{TFI}(c) \equiv \mathcal{TFI}(r)$. \\
    Representative Cone (\textit{RC}) 
    & A subgraph within $\mathcal{TFI}(r)$, rooted at the representative node $r$, $\mathcal{RC}(r) \subseteq \mathcal{TFI}(r)$ with $r$ as the unique sink. \\
    
    Choice Cone (\textit{CC}) 
    & A structural variant of $\mathcal{RC}(r)$, rooted at choice node $c$ in the same equivalence class ($c \sim r$), sharing leaf nodes with $\mathcal{RC}(r)$ but diverging in internal structure. $\mathcal{CC}(c) \equiv \mathcal{RC}(r)$ with $c$ as the unique sink. \\

    \bottomrule
  \end{tabular}
  \label{tab:definitions}
\end{table*}
We begin by summarizing several key terms that will be  used in the paper. A detailed terminology table is provided in \Cref{terminology} to facilitate understanding.

\subsection{And-Inverter Graphs (AIG) with Choice}
An AIG, denoted as $\mathcal{G}$, is a directed acyclic graph (DAG) that represents Boolean functions. Each node in $\mathcal{G}$ is either a \textit{primary input} (PI), which has no incoming edges, or a two-input AND gate with exactly two incoming edges. Additionally, some nodes in $\mathcal{G}$ are designated as \textit{primary outputs} (POs), which define the graph’s outputs. Edges in $\mathcal{G}$ can optionally indicate signal inversion.
\begin{figure}[!t]
    \centering
    \includegraphics[width=0.45\textwidth]{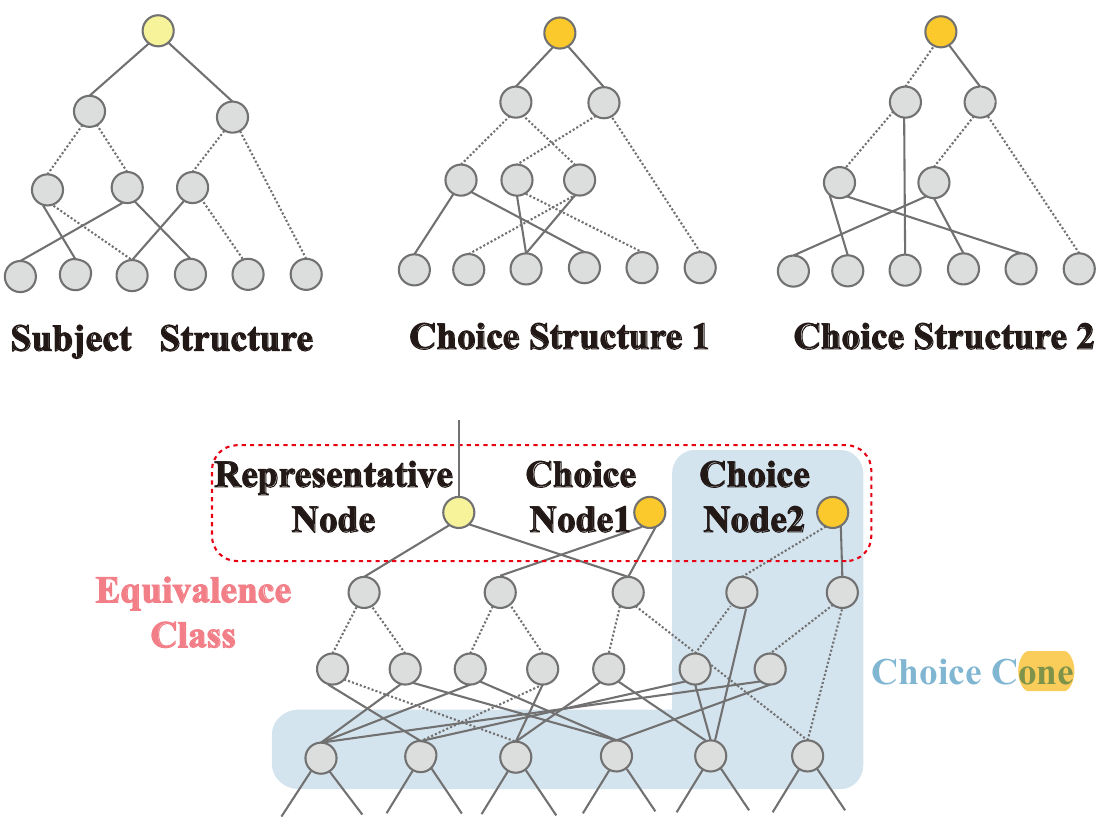}
    \caption{AIG with Choices.}
    \label{fig:choice}
\end{figure}

Functional equivalence relation $\sim$ is defined on the set of nodes $\mathcal{X}$ in $\mathcal{G}$. Two nodes $n_1, n_2 \in \mathcal{X}$ are considered equivalent, i.e., $n_1 \sim n_2$, if they implement the same Boolean function, such that $f(n_1) = f(n_2)$. Equivalence classes maintain the relationships of all equivalent nodes in $\mathcal{G}$. Each equivalence class $C \in \mathcal{X}$ is a set of nodes ${n_1, n_2, n_3, \dots}$, where every pair of nodes within the class is functionally equivalent, i.e., $f(n_1) = f(n_2) = f(n_3) = \dots$.

An \textit{AIG with choice} as shown in~\Cref{fig:choice}~\cite{chatterjee2007algorithms} is an extension of an AIG with equivalence classes. For each equivalence class, the first node in the class is the Representative Node from the subject graph and is designated as the \textit{functional representative}, which must have outgoing edges. The remaining nodes in the equivalence class are called \textit{choice nodes}, which are dangling nodes with no outgoing edges. 


\begin{figure}[h]
    \flushright 
    \includegraphics[width=0.5\textwidth]{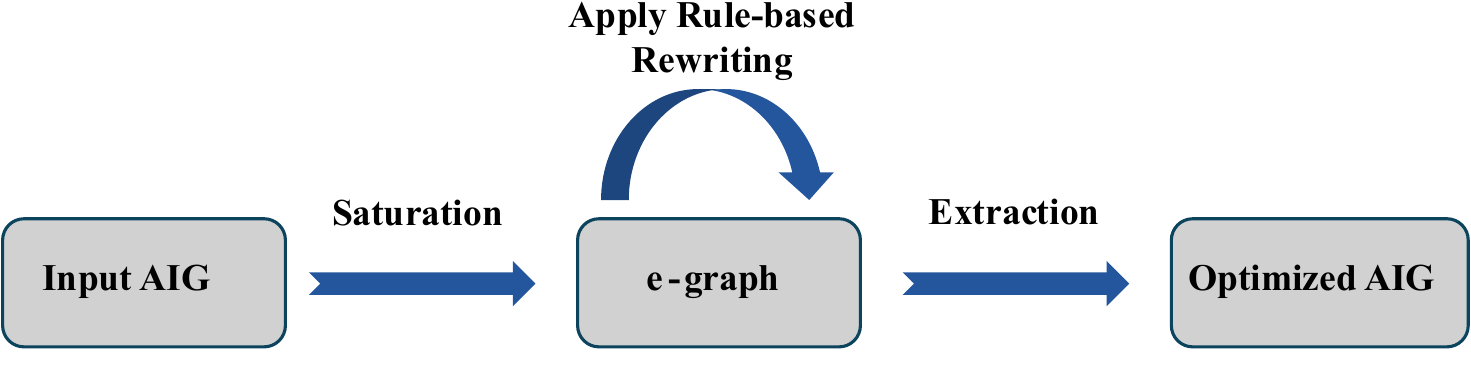}
    \captionsetup{justification=centering} 
    \caption{Equality Saturation Workflow }
    \label{fig:egraph}
\end{figure}
\subsection{Technology Mapping for AIG with Choice in Standard Cells}
Technology mapping~\cite{chatterjee2007algorithms}~\cite{cong} translates Boolean networks into either standard cell implementations (matching to library gates) or FPGA architectures (synthesizing via configurable k-LUTs), optimizing area/delay metrics under timing constraints. For standard cell mapping, it is performed through three coordinated phases. 1) $k$-feasible cuts~\cite{cong1999cut} are computed for functional representatives of equivalence classes in $\mathcal{G}$ with choice. 2) These cuts are then matched to library cells via Boolean matching~\cite{booleanmatching}. 3) an optimal cover is generated by selecting a non-overlapping set of matched cuts that minimizes either the critical path delay $\Delta_{\text{crit}}$ for timing-driven designs or the total area $\sum \text{area}(c)$ for area-constrained implementations. For details on mapping, we refer the reader to~\cite{chatterjee2007algorithms}.

\subsection{Structural Bias Issues in Technology Mapping}
Structural bias~\cite{chatterjee2006reducing} arises in technology mapping when the final mapped implementation becomes overly dependent on the input Boolean network’s structure due to the local nature of the matching process.  While Boolean matching mitigates local bias by enabling functional matching instead of structure pattern matching through supergates and multi-phase compatibility, allowing diverse implementations of the same logic, it cannot resolve global bias because these cuts are chosen sequentially based on a heuristic that optimizes for local area cost, creating path dependencies that lock in the global structure.

\subsection{Equality Saturation for Boolean Network}

Equality saturation~\cite{egg} has recently become a powerful optimization technique, widely used in Boolean optimization~\cite{ustun2022impress}~\cite{chen2024syn}~\cite{chen2025emorphicscalableequalitysaturation}~\cite{coward2022automatic}~\cite{coward2023automating}, compiler optimization~\cite{cheng2024seer}~\cite{cai2025smoothe}, and formal verification~\cite{yin2025boole}~\cite{coward2023datapath}. It operates in two phases: saturation and extraction as shown in~\Cref{fig:egraph}. In the saturation phase, rewriting rules are iteratively applied for a fixed number of iterations or saturation. This process generates a large number of equivalent expressions and stores them in a compact data structure called an e-graph, which represents equivalence classes of terms while maintaining congruence relations. In the extraction phase, an optimized term is selected from the e-graph using a cost function that evaluates the entire graph to determine the most optimal term. This concise, two-phase approach not only makes equality saturation an effective and non-destructive optimization method but also enables the exploration of a vast space of equivalent structures.  Related to equality saturation, we refer readers to \cite{egg} and Section II of \cite{chen2025emorphicscalableequalitysaturation}.

\begin{figure*}[!t]
–    \centering
    \begin{subfigure}[b]{0.45\textwidth}
        \centering
        \includegraphics[width=\textwidth]{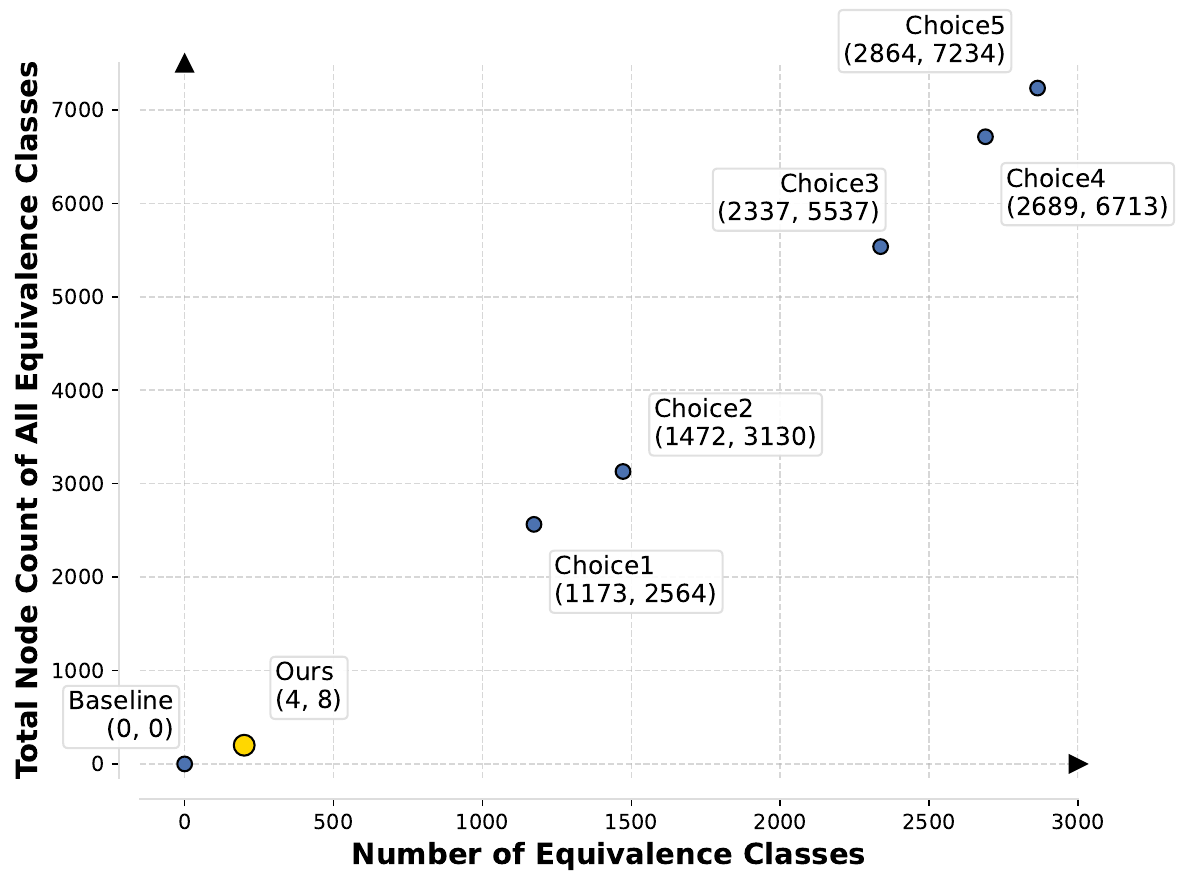}
        \caption{Total Choice Nodes per Case.}
        \label{fig:barplot}
    \end{subfigure}
    \hfill
    \begin{subfigure}[b]{0.45\textwidth}
        \centering
        \includegraphics[width=\textwidth]{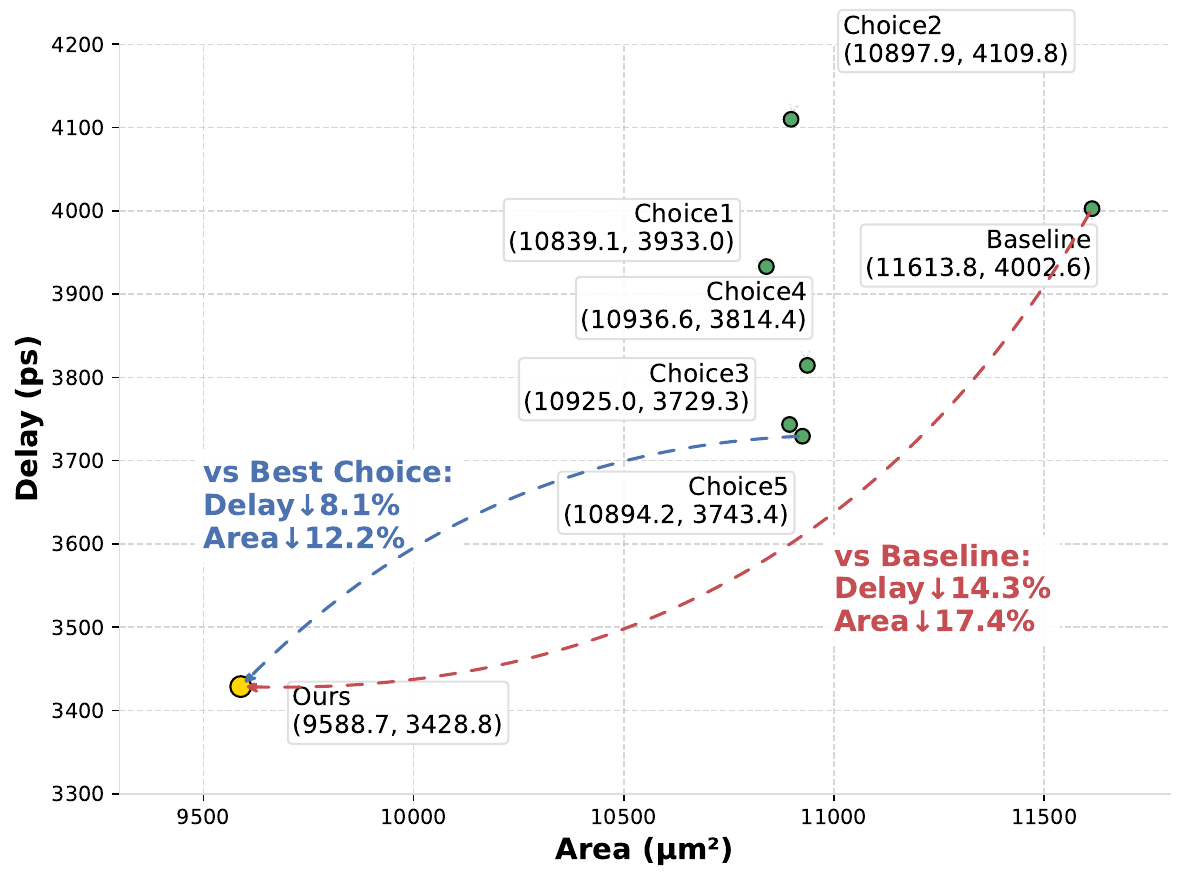}
        \caption{Delay vs Area comparison across methods.}
        \label{fig:scatter}
    \end{subfigure}
    \caption{Rethinking Choice Exploration: The Impact of Quantity and Quality of Choices on Technology Mapping.}
    \label{fig:motivation_combined}
\end{figure*}
\section{RELATED WORK and MOTIVATION}

In this section, we review related works that address the structural bias problem and the construction of choice networks. Mailhot and De Micheli proposed a Boolean matching~\cite{booleanmatching} approach that eliminates local structural bias by directly matching a gate with a subgraph through the comparison of their Boolean functions, rather than relying solely on structural pattern matching. Chen and Cong leveraged Binary Decision Diagrams (BDDs) to detect choices~\cite{cong}; however, the computational complexity of BDDs limits their scalability to small designs. Lehman introduced the use of different algebraic decompositions to construct local mapping graphs~\cite{lehman1997logic}, but this approach primarily provides local choices. Another notable method combines random simulation and SAT-solving for choice detection, which has been demonstrated to be both robust and scalable in the combinational equivalence checking literature~\cite{history}~\cite{sat}. The concept of a choice network for logic synthesis was introduced to address structural bias by integrating multiple networks into an AIG with choices~\cite{chatterjee2006reducing}~\cite{chatterjee2007algorithms}~\cite{mishchenko2006improvements}, providing a scalable and efficient method for generating choices.

Recently LCH~\cite{grosnit2023lightweight} extracts non-overlapping subgraphs to construct a choice network with lighter-weight choices and achieves faster processing speeds. However, it does not analyze the impact of the generated choices on mapping quality. MCH~\cite{hu2025mixedstructuralchoiceoperator} proposes a method that leverages a Mixed Structural Choice Operator to incorporate heterogeneous representations, such as XMG~\cite{XIG}, MIG~\cite{MIG}, and AIG, into a mixed choice network. Nevertheless, it focuses on smaller-scale local choice cones and using local choice mapping results for choice selections. E-morphic~\cite{chen2025emorphicscalableequalitysaturation} provides a method for global resynthesis using equality saturation and simulated annealing that adjusts subject graph structures based on mapping results after independent optimization. However, this approach faces runtime and efficiency limitations in large-scale cases as it requires full-circuit mapping to obtain cost feedback.

Motivated by the limitations of prior approaches, we present a case study contrasting our method with lossless synthesis for constructing choice networks in \texttt{ABC}, demonstrating how we alleviate recent methodological constraints. As shown in~\Cref{fig:motivation_combined}, \textit{Choice1} represents the vanilla implementation of the Boolean choice network construction in \texttt{ABC}. We enhance \texttt{ABC}'s vanilla implementation by increasing the number of intermediate optimization phases, which preserve historical networks across iterative synthesis steps, thereby constructing \textit{Choice2-5} configurations with expanded choice sets, using experimental data from the arithmetic design "sin" in the EPFL~\cite{amaru2015epfl} benchmarks. In~\Cref{fig:barplot}, the x-axis denotes the number of equivalence classes identified in each choice network, while the y-axis quantifies the total node count within all equivalence classes, calculated as the sum of choice nodes and representative nodes. From the ~\Cref{fig:scatter}, we demonstrate the Quality of Results (QoR) for technology mapping across different choice networks constructed in \texttt{ABC} and our \deftitle method. The yellow points mark the Representative Cones (\textit{RCs}) chosen by our \deftitle method, which strategically filters impact‑critical \textit{RCs} and synthesizes higher-quality Choice Cones \textit{(CCs)} achieves an 8.1\% delay reduction and a 12.2\% area improvement after post mapping, outperforming the best-performing 5-choice networks. 

\textbf{Through our case study, we highlight the following key observations:
}

\textbf{1. Localized nature of Choice Cones in lossless synthesis methods:} We analyzed all 2,564 \textit{CCs} extracted from 1,472 equivalence classes in the Choice1 configuration. Notably, over 99.36\% of these cones are subgraphs with fewer than 30 nodes.  
This strong skew toward small cone sizes suggests a potential limitation in the diversity of structural contexts captured, prompting further exploration into whether such localized structures can fully support globally-aware optimization.

\textbf{2. More choices do not necessarily imply better QoR:} 
Our experimental results reveal that increasing the number of choices, such as those generated by the \texttt{dch} command in \texttt{ABC}, does not directly lead to better technology mapping results. 
This observation underscores the importance of carefully selecting Representative Cones \textit{(RCs)} and strategically synthesizing higher-quality Choice Cones \textit{(CCs)} in \deftitle,  rather than relying solely on the quantity of choices.


\section{Methodology}\label{sec:our-method}
\begin{figure*}[!ht]
    \centering
    \includegraphics[width=\textwidth]{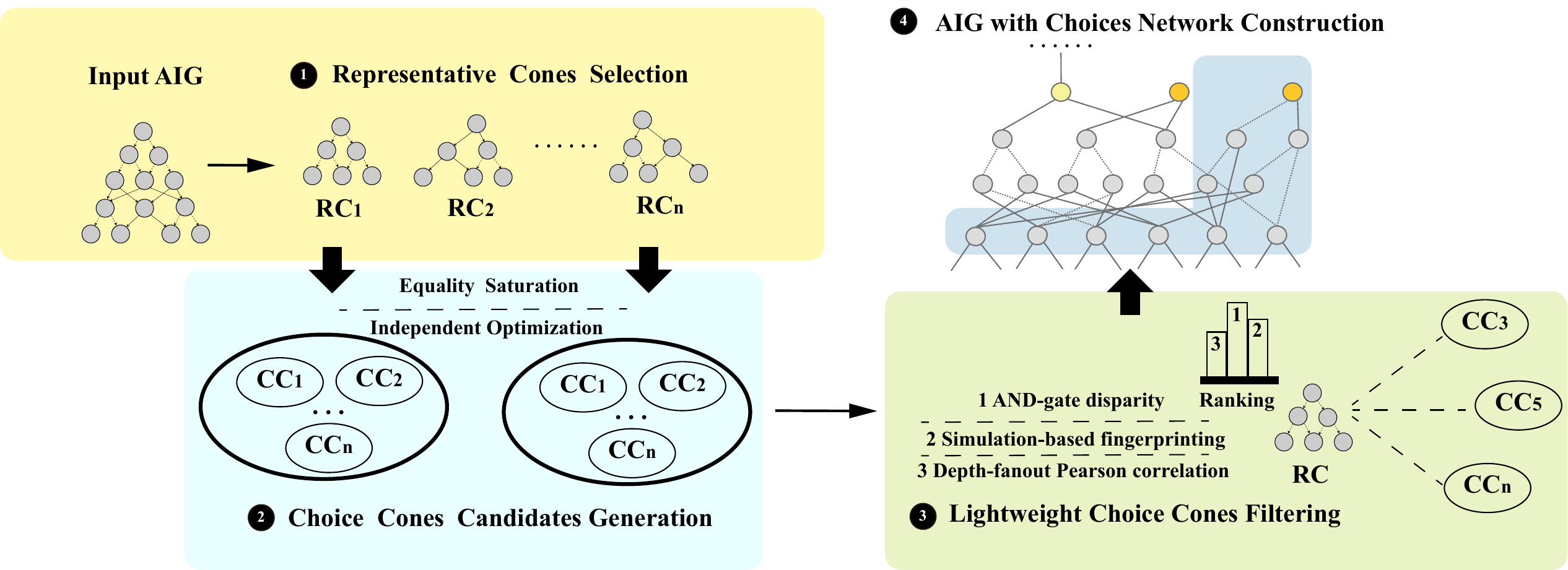}
    \caption{Overview of CRISTAL.}
    \label{fig:overview}
\end{figure*}
\subsection{ Overview of \deftitle}
\Cref{fig:overview} outlines the \deftitle workflow. First, we screen and prune Representative Cones (\textit{RCs}) to retain a minimal set with dominant technology mapping influence (Section~\Cref{method1}).
Choice Cones (\textit{CCs}) synthesis employs a hybrid optimization strategy that combines equality saturation-based structural rewrites with traditional independent logic optimization-based rewriting, enabling the construction of structurally richer and higher-quality choice cones, as illustrated in~\Cref{method2}. For \textit{CCs} filtering, we combine simulation-based fingerprinting, AND-gate disparity, depth-fanout Pearson correlation, and independent AIG metrics (e.g., size and level) to evaluate structural diversity and implementation quality. Normalized hybrid scores are used to rank and filter \textit{CCs} candidates, prioritizing high-impact variants. Finally, \Cref{method4} demonstrates the details of choice network construction and validation, where invalid choices are removed to ensure the functional correctness required for technology mapping.

\subsection{ Representative Cones Selection} \label{method1}

Maximum Fanout-Free Cones (\textit{MFFCs}) and low fanout cones enhance area-oriented technology mapping by enabling localized optimization. \textit{MFFCs} isolates logic regions without external fanout dependencies, allowing independent computation of optimal area-minimizing cuts per subgraph, which reduces redundant duplication. 

We first consider \textit{MFFCs} and low--fanout cones as candidate \textit{RCs} and operate in two modes, \emph{delay} and \emph{area}, as described in Algorithm~\ref{alg:choice_cone_gen}. In the \emph{delay mode}, we select \textit{MFFCs} whose Representative Node~\textit{RC} are on the critical path; these cones also significantly influence both area and delay mapping because their isolation from external fanout dependencies enables simultaneous timing closure and area minimization within the same subgraph. In \emph{area mode}, after performing extensive independent area-oriented optimizations that may eliminate pure \textit{MFFCs}—where all fanouts of every internal node are exclusively contained within the cone itself, ensuring no external connection dependencies—the algorithm adapts by relaxing the \textit{MFFCs} collection criteria. It transitions to enumerating \textit{low-fanout cones}, which are constructed by recursively including fan-in nodes whose global fanout count does not exceed a predefined threshold~$T$. \textit{RCs} are then filtered by a size threshold~$k_{\min}$, prioritizing larger \textit{MFFCs}. Critical larger cones inherently exhibit richer structural diversity, enabling impactful area–delay tradeoffs during mapping, whereas smaller cones often lack meaningful structural variants and contribute minimally to optimization gains. By excluding overly small \textit{RCs}, the approach mitigates runtime inefficiencies caused by excessive enumeration of low-reward candidates, while maintaining focus on critical logic regions that drive significant quality improvements. 
\begin{algorithm}[htbp]
  \caption{Representative Cones Generation}
  \label{alg:choice_cone_gen}
  \SetAlgoLined
  \KwIn{AIG $G$, Mode $M$, Threshold $k$, Fanout limit $T$}
  \KwOut{Set of Representative cones $\mathcal{C}$}

  $\mathcal{C} \gets \emptyset$\;

  $\mathcal{P} \gets 
    \begin{cases}
      \{n \in G \mid \textsc{CriticalPath}(n)\}, & \text{if } M = \textsc{delay} \\
      \{n \in G \mid \textsc{IsAndNode}(n)\}, & \text{otherwise}
    \end{cases}$\;

  \For{each $n \in \mathcal{P}$}{
    $(v_c, v_s) \gets \textsc{MffcConeSupp}(n)$\;
    \If{$|v_s| \geq 2$ \textbf{and} $|v_c| \geq k$}{
      $\mathcal{C} \gets \mathcal{C} \cup \{\textsc{Build\_MFFC}(n)\}$\;
    }
  }

  \If{$\mathcal{C} = \emptyset$}{
    \For{each $n \in \mathcal{P}$}{
      $(v'_c, v'_s) \gets \textsc{LowfanoutConeSupp}(n, T)$\;
      \If{$|v'_s| \geq 2$ \textbf{and} $|v'_c| \geq k$}{
        $\mathcal{C} \gets \mathcal{C} \cup \{\textsc{Build\_Cone}(n, v'_c, v'_s)\}$\;
      }
    }
  }
\end{algorithm}

 \subsection{Diversifying Choice Cones via Hybrid Structural Mutation}  \label{method2}



\begin{figure}[h]
    \flushright 
    \includegraphics[width=0.45\textwidth]{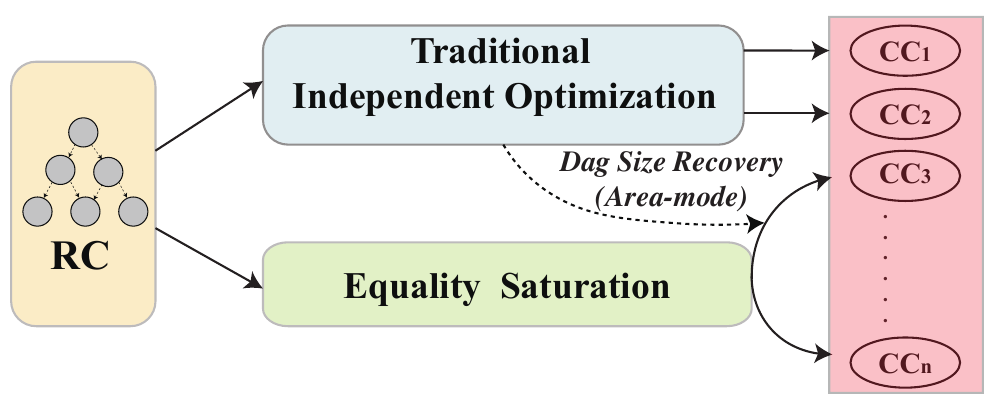}
    \captionsetup{justification=centering} 
    \caption{Generation of Structurally Diverse Choice Cones  }
    \label{fig:eqsat}
\end{figure}

After the Representative Cones (\textit{RCs}) are selected in~\Cref{method1}, we generate a set of structural variants for each \textit{RC}, which serve as \textit{Choice Cones (\textit{CCs})} candidates for following~\Cref{method3}. This step is essential for addressing the global-awareness bias in structural optimization, as it introduces a broader and more diverse set of implementation options for each \textit{RC} as shown in~\Cref{fig:eqsat}.
 Recent work~\cite{chen2025emorphicscalableequalitysaturation} has demonstrated that equality saturation is highly effective for resynthesis, enabling extensive structural exploration guided by technology mapping costs. Building upon this idea, we extend the use of equality saturation to enrich the structural diversity of \textit{CCs}. Our method systematically generates numerous structural variants of each \textit{RC} while preserving independent optimization objectives. In addition to equality‐saturation–based rewrites, We also incorporate the independent optimizations provided by traditional cut-based techniques~\cite{aig2006}~\cite{mishchenko2011delay}, forming a truly hybrid variant-construction pipeline.

\textbf{Implementation Details.} We first convert the set of \textit{RCs} into e-graphs using a custom AIG-to-e-graph parser. This representation compactly encodes structural equivalence classes, enabling efficient exploration of Boolean logic transformations. A series of boolean rewriting rules (as detailed in~\cite{chen2024syn}) are applied to saturate the e-graph — a compact data structure that maintains a large set of functionally equivalent but structurally diverse representations. E-graph extraction involves selecting the best-cost enode from each equivalence class to form a graph equivalent to the input graph. For cost-aware extraction, we assign costs to each \textit{enode} in each equivalence class using a bottom-up, greedy heuristic based on Abstract Syntax Tree (AST) depth and AST size. To reduce extraction complexity, only enodes with costs $\leq$ the minimum cost in their equivalence classes are retained. This strategy not only accelerates extraction by reducing the search space but also preserves structural diversity. 
 Notably, tree-based cost models are particularly well-suited for \emph{delay mode} optimization, as delay estimation naturally follows the bottom-up evaluation order of ASTs and avoids complications from shared substructures.
 While DAG-based cost models could theoretically improve DAG-size optimization, they face NP-complete extraction complexity and risk cyclic dependencies.
 Instead, in the \emph{area mode}, to better account for practical usage and scalability, instead of aiming to maintain DAG-size optimality, we use tree-based area cost models to construct alternative structures. These structures are further refined by integrating two passes of the \textit{compress2rs} operator from \texttt{ABC} to perform area recovery after converting the e-graph back to AIG.
 Finally, we describe our complete CCs candidate‑generation workflow: For each \textit{RC}, we first generate some CCs supported by \texttt{ABC} using \texttt{mfs} and \texttt{compress2rs} in the \emph{area mode}, and \texttt{if -g} in the \emph{delay mode}. These are then combined with the structural variants generated above using equality saturation. This hybrid approach produces a larger set of structural variants, forming a highly diverse set of \textit{CCs} for each \textit{RC}.

\subsection{Lightweight Priority-Ranked Choice Cones Filtering}
\label{method3}
 Within each equivalence class, lightweight prioritization of the \textit{RC}’s \textit{CCs} candidates is critical for balancing mapping quality and runtime efficiency. While existing methods like graph edit distance~\cite{gao2010survey} (NP-hard) can measure structural similarity, their prohibitive computational costs make them impractical for large-scale practical application. To address this challenge, we develop a priority-ranking evaluation method that integrates three lightweight structural metrics with AIG-specific quality (e.g., logic depth and node count),  enabling efficient yet comprehensive assessment of structural diversity and independent quality. 
\subsubsection{Problem Formulation} 
For each equivalence class, let $\mathcal{G} = \{G_1, G_2, \dots, G_N\}$ be the set of \textit{CCs} generated via~\Cref{method2} from the \textit{RC} $G_0$. Both the \textit{RC} and \textit{CCs} are represented as AIG. Our objective is to select the optimal candidate $G^* \in \mathcal{G}$ that balances structural diversity and implementation quality relative to \textit{RC} $G_0$:

\begin{equation}
G^* = \argmax_{G \in \mathcal{G}}
  \bigl(\alpha\,S_{\mathrm{hybrid}}(G) + \beta\,Q_{\mathrm{obj}}(G)\bigr).
\end{equation}

where $S_{\text{hybrid}}$ represents the structural diversity score, $Q_{\text{obj}}$ denotes the implementation quality (level/size), and $\alpha,\beta$ are Weighting coefficients controlling diversity-quality tradeoff.

\subsubsection{Efficient Hybrid Structural Similarity Evaluation Model}
\label{subsec:hybrid_eval}

Our evaluation model combines three orthogonal lightweight metrics%
—simulation-based fingerprinting~($S_{\mathrm{sim}}$), AND‑gate disparity~($S_{\mathrm{and}}$), 
and depth–fanout Pearson correlation~($S_{\mathrm{pearson}}$)%
—for efficient structural dissimilarity assessment. Each metric is first 
normalized, and the final hybrid score is computed as
\[
S_{\mathrm{hybrid}}
= S_{\mathrm{sim}} + S_{\mathrm{and}} + S_{\mathrm{pearson}}.
\]

\paragraph{Simulation‑Based Fingerprinting:} This algorithm evaluates the structural dissimilarity between two AIG by combining bit-parallel simulation and logical fingerprint comparison. First, \(64 \times n_{\text{words}}\) random input patterns are generated and propagated through both the \textit{RC} AIG (\(G_{\text{subj}}\)) and the candidate \textit{CC} AIG (\(G_{\text{cand}}\)) in topological order, producing \(n_{\text{words}}\) 64-bit simulation vectors per node. To ensure invariance to inverter reordering — we normalize \( G_{\text{subj}} \)'s vectors using the least significant bit (LSB) of the first 64-bit word as a polarity indicator: if LSB = 1, the entire vector is inverted
 (\(\mathbf{s}(v) \gets \neg \mathbf{s}(v)\)). 
 The normalized vectors are stored in a hash table \(H_r\) as unique logical fingerprints. For \(G_{\text{cand}}\), the simulation vector of each node is checked against the hash table, both in its original form and its bitwise complement, to account for potential polarity differences. A node in \(G_{\text{cand}}\) is marked as a mismatch if neither its vector nor its complement exists in \(H_r\). Finally, the algorithm computes a normalized mismatch score (\(S\)) as the proportion of unmatched nodes in \(G_{\text{cand}}\), capturing the structural divergence between the two graphs. This fingerprinting approach enables rapid and scalable detection of both structural dissimilarity and logical mismatches, offering a highly efficient yet sensitive measure of structural variation.

\begin{algorithm}[t]
  \caption{Simulation-based Structure Dissimilarity}
  \label{alg:structure_dissimilarity}
  \SetAlgoLined
  \KwIn{Subject graph $G_{\text{subj}}$, candidate graph $G_{\text{cand}}$, word count $n_{\text{words}}$}
  \KwOut{Dissimilarity score $S$}
  
  Generate $P \gets \{\phi_1, \phi_2, \dots, \phi_{64 \times n_{\text{words}}}\}$ random patterns\;
  Simulate $P$ through $G_{\text{subj}}$ and $G_{\text{cand}}$ in topological order\;
  
  \For{each node $v \in G_{\text{subj}}$}{
    Let $\mathbf{s}_v \in \{0,1\}^{64 \times n_{\text{words}}}$ be the simulation vector of $v$\;
    $\text{compl} \gets \mathbf{s}_v[0]$\;
    \If{$\text{compl} = 1$}{
      $\mathbf{s}_v \gets \neg \mathbf{s}_v$\;
    }
    Insert $\mathbf{s}_v$ into hash table $H_r$ for $G_{\text{subj}}$\;
  }
  
  \For{each node $u \in G_{\text{cand}}$}{
    Let $\mathbf{s}_u \in \{0,1\}^{64 \times n_{\text{words}}}$ be the simulation vector of $u$\;
    \If{$\mathbf{s}_u \notin H_r$ \textbf{and} $\neg \mathbf{s}_u \notin H_r$}{
      Mark $u$ as a mismatch\;
    }
  }

  \Return $S = \frac{\text{\# mismatches}}{|G_{\text{cand}}|}$\;
\end{algorithm}

\paragraph{Depth-Fanout Pearson Correlation}

The topological similarity analysis employs Pearson correlation coefficient~\cite{cohen2009pearson} for depth/fanout distribution matching: This method uses the Pearson correlation coefficient \(r \in [-1, 1]\), which quantifies the linear relationship between two datasets: \(r = 1\) indicates perfect positive correlation, \(r = 0\) indicates no correlation, and \(r = -1\) indicates perfect inverse correlation. To quantify structural dissimilarity, the mismatch score is defined as \(S_{\text{pearson}} = 1 - |r|\), where higher values indicate greater structural divergence. The algorithm traverses the \textit{RC} (\(G_{\text{subj}}\)) and the candidate \textit{CC} (\(G_{\text{cand}}\)) to compute the depth and fanout values for each node, generating two sequences \(d_{\text{subj}}, d_{\text{cand}}\) for depth and \(f_{\text{subj}}, f_{\text{cand}}\) for fanout. The number of nodes included in the computation is limited by \(m = \min(|G_{\text{subj}}|, |G_{\text{cand}}|)\), ensuring statistical fairness by comparing only the overlapping node range without artificial padding, which avoids distorting distribution patterns and focuses on core structural comparability. The Pearson coefficients for depth and fanout sequences are then computed, reflecting the global structural alignment. This truncation strategy guarantees valid pairwise comparisons while minimizing bias from unmatched graph scales. The final dissimilarity score \(S_{\text{pearson}}\) links global depth and fanout patterns to structural similarity.

By integrating the aforementioned three structural diversity measurements with independent metrics, the priority-ranking mechanism utilizes normalized hybrid scores ($S_\text{hybrid}$) to filter and rank \textit{CC} candidates, selecting the top-3 most optimal \textit{CCs} for each \textit{RC} to advance to technology mapping, thereby reducing redundant choices and mitigating their impact on area-oriented heuristics (e.g., area-flow), as an excessive number of \textit{CCs} per \textit{RC} can introduce many node fanouts that do not belong to the original subject graph, which in turn reduces the accuracy of heuristic area estimation. This also reduces the burden of Boolean matching and improves the mapping efficiency.
\begin{algorithm}[t]
  \caption{Depth-Fanout Pearson Correlation}
  \label{alg:pearson_correlation}
  \SetAlgoLined
  \KwIn{Subject graph $G_{\text{subj}}$, candidate graph $G_{\text{cand}}$}
  \KwOut{Structural dissimilarity score $S_{\text{pearson}}$}

  \For{each node $v_i$ in both $G_{\text{subj}}$ and $G_{\text{cand}}$}{
    Compute depth and fanout: $d_{\text{subj}}[i]$, $f_{\text{subj}}[i]$, $d_{\text{cand}}[i]$, $f_{\text{cand}}[i]$\;
  }

  Let $m = \min(|V_{\text{subj}}|, |V_{\text{cand}}|)$\;
  Compute Pearson correlation coefficient:
  \[
  r = \frac{\sum_{i=1}^m (d_{\text{subj}}[i] - \bar{d}_{\text{subj}})(d_{\text{cand}}[i] - \bar{d}_{\text{cand}})}{\sqrt{\sum_{i=1}^m (d_{\text{subj}}[i] - \bar{d}_{\text{subj}})^2} \cdot \sqrt{\sum_{i=1}^m (d_{\text{cand}}[i] - \bar{d}_{\text{cand}})^2}}
  \]

  Compute structural dissimilarity score:
  \[
  S_{\text{pearson}} = 1 - |r|
  \]

  \Return $S_{\text{pearson}}$\;
\end{algorithm}

\subsection{Construction and Validation of Choice Networks :}
\label{method4}
For each \textit{RC}, the corresponding \textit{CCs} filtered via~\Cref{method3} are integrated into the subject graph through a choice network, where each equivalence class aggregates functionally equivalent but structurally distinct variants. During construction, all \textit{CCs} are merged into the AIG via structural hashing, with their corresponding root choice nodes temporarily marked as dangling nodes (nodes without fanout). These root choice nodes are then linked to their representative node through linked lists to form equivalence classes. To ensure validity, two critical checks are enforced (Algorithm~\ref{alg:remove_bad_choices}): (1) \textit{Acyclicity verification} detects and removes cyclic dependencies in the $\mathcal{T}$-relation (transitive dependency closure) via depth-first search, as cycles disrupt the topological ordering required for cut enumeration in technology mapping~\cite{chatterjee2007algorithms}; (2) \textit{Fanout-free validation} eliminates any \textit{CC} whose root choice node exhibits fanout connections—a condition that may arise during structural hashing when integrating \textit{CCs} into the subject graph. Overlaps between \textit{CCs} or between a \textit{CC} and its \textit{RC} can lead to logic subsumption, where structural hashing inadvertently merges portions of the AIG. For example, if \textit{CC} B is entirely subsumed by \textit{RC} A during hashing, the root node of \textit{CC} 
B—originally a dangling node—may inherit fanout edges from \textit{RC} A's internal structure. These restrictions are fundamental for cut enumeration and Boolean matching. By enforcing strict acyclicity and fanout isolation, the choice network guarantees compatibility with technology mapping algorithms.
\begin{algorithm}[h]
  \caption{Remove Bad Choices}
  \label{alg:remove_bad_choices}
  \SetAlgoLined
  \KwIn{AIG $G$ with equivalence classes}
  \KwOut{AIG $G$ with bad choices removed}
  
  Mark fanout for all nodes in $G$\;
  
  \For{each equivalence class $\mathcal{C}$ in $G$}{
    Let $prev \gets$ representative node of $\mathcal{C}$\;
    \For{each node $curr \in \mathcal{C} \setminus \{prev\}$}{
      \If{$curr$ has fanout \textbf{or} TFI check fails for (representative node, $curr$)}{
        Remove $curr$ from $\mathcal{C}$\;
      }
      \Else{
        $prev \gets curr$\;
      }
    }
  }
  
  \Return $G$\;
\end{algorithm}

\section{Experiments}\label{sec:experiment}

\subsection{Experimental Setup} 

All experiments were conducted on a server running Ubuntu 20.04.4 LTS, equipped with Intel\textregistered{} Xeon\textregistered{} Gold 6418H processors and 64 GB of memory. The \deftitle framework was implemented based on the open-source logic synthesis tool \texttt{ABC}. For choice generation using equality saturation, we interfaced with the existing \texttt{egg} library in Rust for e-graphs. The conversions between AIG and e-graphs (AIG-to-e-graph and e-graph-to-AIG) were implemented using \texttt{Flussab}~\footnote{https://github.com/jix/flussab}, a Rust-based AIG parser, with modifications tailored to our specific use case. The test circuits used in the evaluation include datapath and control path circuits from the IWLS2005~\footnote{https://iwls.org/iwls2005/benchmarks.html}, EPFL~\cite{amaru2015epfl}, and ISCAS89~\cite{Brglez1989} benchmarks. The mapping quality was evaluated using the ASAP7 technology library~\cite{clark2016asap7} . All results were formally verified using \texttt{ABC}’s \texttt{cec} command to ensure correctness. The algorithm selects \textit{RCs} using size thresholds {800, 85, 30, 20, 15, 10}, prioritizing \textit{MFFCs}. If a valid \textit{MFFC} cannot be identified, the algorithm relaxes the selection criteria. It then attempts to construct a \textit{low-fanout cone} by recursively including fanin nodes whose global fanout count is bounded by a threshold ($K=3$). To mitigate computational overhead, a cardinality constraint limits \textit{RCs} per threshold to 10 if candidate \textit{RCs} exceed 100—common for small cones (e.g., \( \leq 20 \ \text{nodes} \)), which offer limited structural diversity. We set $\alpha:\beta = 2:3$ for the diversity–quality tradeoff. For each \textit{RC}, we select the top 3 \textit{CCs} according to~\Cref{method3} from 10 \textit{CCs} generated by~\Cref{method2}. To demonstrate the performance of \deftitle, we prepared four research questions:
\begin{table*}[!t]
    \caption{QoR and runtime comparison between \deftitle and the \texttt{dch} operator in \texttt{ABC}.}
    \centering
    \resizebox{\linewidth}{!}{%
    \begin{tabular}{l|rrr|rrr|rrr|rrr|rrr|rrr|}
    \hline
\multicolumn{1}{c|}{\multirow{3}{*}{Circuit}}
  & \multicolumn{9}{c|}{\textbf{Area-Oriented}}
  & \multicolumn{9}{c|}{\textbf{Delay-Oriented}} \\
\cline{2-19}
  & \multicolumn{3}{c|}{\textbf{\textit{dch}}}
  & \multicolumn{3}{c|}{\textbf{\textit{\deftitle\ w Priority-rank}}}
  & \multicolumn{3}{c|}{\textbf{\textit{\deftitle\ w/o Priority-rank}}}
  & \multicolumn{3}{c|}{\textbf{\textit{dch}}}
  & \multicolumn{3}{c|}{\textbf{\textit{\deftitle\ w Priority-rank}}}
  & \multicolumn{3}{c|}{\textbf{\textit{\deftitle\ w/o Priority-rank}}} \\
\cline{2-19}
     &
    \multicolumn{1}{c}{Area ($\mu m^2$)} & 
    \multicolumn{1}{c}{Delay ($ps$)} &  
    \multicolumn{1}{c|}{$\mathrm{RT}$($s$)} & 
    \multicolumn{1}{c}{Area ($\mu m^2$)} & 
    \multicolumn{1}{c}{Delay ($ps$)} & 
    \multicolumn{1}{c|}{$\mathrm{RT}$ ($s$)}&
        \multicolumn{1}{c}{Area ($\mu m^2$)} & 
    \multicolumn{1}{c}{Delay ($ps$)} & 
    \multicolumn{1}{c|}{$\mathrm{RT}$ ($s$)}&
    \multicolumn{1}{c}{Area ($\mu m^2$)} & 
    \multicolumn{1}{c}{Delay ($ps$)} &  
    \multicolumn{1}{c|}{$\mathrm{RT}$ ($s$)} & 
    \multicolumn{1}{c}{Area ($\mu m^2$)} & 
    \multicolumn{1}{c}{Delay ($ps$)} & 
    \multicolumn{1}{c|}{$\mathrm{RT}$($s$)}&
        \multicolumn{1}{c}{Area ($\mu m^2$)} & 
    \multicolumn{1}{c}{Delay ($ps$)} & 
    \multicolumn{1}{c|}{$\mathrm{RT}$($s$)}
    \\
   \hline
         cavlc & 442.30 & 180.55 & 1 & 448.13 & \cellcolor[rgb]{0.77, 0.87, 0.71}\textbf{174.75} & 7 & \cellcolor[rgb]{0.77, 0.87, 0.71}\textbf{440.90} & \cellcolor[rgb]{0.77, 0.87, 0.71}\textbf{172.98} & 6 & 504.12 & 98.79 & 1 & 512.75 & 101.31 & 7 & 508.32 & 105.60 & 7 \\ \hline
        div & 20227.71 & 68173.25 & 8 & 20522.81 & \cellcolor[rgb]{0.77, 0.87, 0.71}\textbf{67436.80} & 11 & 20657.88 & \cellcolor[rgb]{0.77, 0.87, 0.71}\textbf{67594.58} & 11 & 68609.75 & 32405.68 & 44 & \cellcolor[rgb]{0.77, 0.87, 0.71}\textbf{65418.71} & \cellcolor[rgb]{0.77, 0.87, 0.71}\textbf{30544.16} & \cellcolor[rgb]{0.77, 0.87, 0.71}\textbf{31} & \cellcolor[rgb]{0.77, 0.87, 0.71}\textbf{66290.24} & \cellcolor[rgb]{0.77, 0.87, 0.71}\textbf{30841.41} & 65 \\ \hline
        Hyp & 195113.77 & 358924.88 & 289 & \cellcolor[rgb]{0.77, 0.87, 0.71}\textbf{194651.86} & \cellcolor[rgb]{0.77, 0.87, 0.71}\textbf{348329.88} & \cellcolor[rgb]{0.77, 0.87, 0.71}\textbf{53} & \cellcolor[rgb]{0.77, 0.87, 0.71}\textbf{194834.98} & \cellcolor[rgb]{0.77, 0.87, 0.71}\textbf{348298.91} & \cellcolor[rgb]{0.77, 0.87, 0.71}\textbf{112} & 300269.62 & 324754.09 & 277 & \cellcolor[rgb]{0.77, 0.87, 0.71}\textbf{298610.53} & \cellcolor[rgb]{0.77, 0.87, 0.71}\textbf{314339.59} & \cellcolor[rgb]{0.77, 0.87, 0.71}\textbf{106} & \cellcolor[rgb]{0.77, 0.87, 0.71}\textbf{289901.97} & \cellcolor[rgb]{0.77, 0.87, 0.71}\textbf{317073.91} & \cellcolor[rgb]{0.77, 0.87, 0.71}\textbf{116} \\ \hline
        i2c & 732.03 & 201.74 & 1 & 747.20 & 209.54 & 6 & 746.03 & 209.54 & 5 & 828.61 & 179.89 & 1 & 875.5 & \cellcolor[rgb]{0.77, 0.87, 0.71}\textbf{146.82} & 7 & 835.14 & 208.46 & 5 \\ \hline
        log2 & 22694.64 & 5914.08 & 16 & \cellcolor[rgb]{0.77, 0.87, 0.71}\textbf{22442.70} & 5978.96 & \cellcolor[rgb]{0.77, 0.87, 0.71}\textbf{13} & 23025.90 & 6036.70 & 16 & 47818.67 & 7337.99 & 33 & \cellcolor[rgb]{0.77, 0.87, 0.71}\textbf{41630.45} & \cellcolor[rgb]{0.77, 0.87, 0.71}\textbf{6490.59} & \cellcolor[rgb]{0.77, 0.87, 0.71}\textbf{15} & \cellcolor[rgb]{0.77, 0.87, 0.71}\textbf{42339.85} & \cellcolor[rgb]{0.77, 0.87, 0.71}\textbf{6447.42} & \cellcolor[rgb]{0.77, 0.87, 0.71}\textbf{19} \\ \hline
        max & 2022.54 & 3118.97 & 1 & 2096.95 & \cellcolor[rgb]{0.77, 0.87, 0.71}\textbf{2939.59} & 9 & 2121.91 & 5050.35 & 25 & 2823.85 & 1368.50 & 1 & 3105.66 & \cellcolor[rgb]{0.77, 0.87, 0.71}\textbf{1192.52} & 8 & 3228.83 & 1440.68 & 8 \\ \hline
        mem\_ctrl & 25894.08 & 1830.25 & 18 & 25895.48 & \cellcolor[rgb]{0.77, 0.87, 0.71}\textbf{1749.78} & \cellcolor[rgb]{0.77, 0.87, 0.71}\textbf{11} & 26067.17 & \cellcolor[rgb]{0.77, 0.87, 0.71}\textbf{1750.01} & \cellcolor[rgb]{0.77, 0.87, 0.71}\textbf{11} & 34051.41 & 1536.63 & 21 & \cellcolor[rgb]{0.77, 0.87, 0.71}\textbf{33424.12} & \cellcolor[rgb]{0.77, 0.87, 0.71}\textbf{1352.29} & \cellcolor[rgb]{0.77, 0.87, 0.71}\textbf{8} & \cellcolor[rgb]{0.77, 0.87, 0.71}\textbf{33357.88} & \cellcolor[rgb]{0.77, 0.87, 0.71}\textbf{1374.93} & \cellcolor[rgb]{0.77, 0.87, 0.71}\textbf{10} \\ \hline
        multiplier & 19812.94 & 4285.56 & 7 & \cellcolor[rgb]{0.77, 0.87, 0.71}\textbf{19335.65} & \cellcolor[rgb]{0.77, 0.87, 0.71}\textbf{4068.79} & 10 & \cellcolor[rgb]{0.77, 0.87, 0.71}\textbf{19445.75} & \cellcolor[rgb]{0.77, 0.87, 0.71}\textbf{4135.06} & 8 & 38059.63 & 2703.45 & 11 & \cellcolor[rgb]{0.77, 0.87, 0.71}\textbf{33716.89} & \cellcolor[rgb]{0.77, 0.87, 0.71}\textbf{2469.20} & \cellcolor[rgb]{0.77, 0.87, 0.71}\textbf{6} & \cellcolor[rgb]{0.77, 0.87, 0.71}\textbf{33688.66} & \cellcolor[rgb]{0.77, 0.87, 0.71}\textbf{2488.26} & \cellcolor[rgb]{0.77, 0.87, 0.71}\textbf{6} \\ \hline
        priority & 362.75 & 628.27 & 1 & \cellcolor[rgb]{0.77, 0.87, 0.71}\textbf{353.42} & \cellcolor[rgb]{0.77, 0.87, 0.71}\textbf{504.97} & 4 & \cellcolor[rgb]{0.77, 0.87, 0.71}\textbf{362.52} & \cellcolor[rgb]{0.77, 0.87, 0.71}\textbf{500.28} & 3 & 723.17 & 725.98 & 1 & 793.85 & 734.11  & 4 & 783.59 & 734.07 & 4 \\ \hline
        router & 113.84 & 196.53 & 1 & \cellcolor[rgb]{0.77, 0.87, 0.71}\textbf{108.94} & \cellcolor[rgb]{0.77, 0.87, 0.71}\textbf{189.96} & 6 & \cellcolor[rgb]{0.77, 0.87, 0.71}\textbf{107.31} & \cellcolor[rgb]{0.77, 0.87, 0.71}\textbf{191.85} & 4 & 310.96 & 253.25 & 1 & \cellcolor[rgb]{0.77, 0.87, 0.71}\textbf{280.4} & \cellcolor[rgb]{0.77, 0.87, 0.71}\textbf{198.75} & 8 & 317.96 & \cellcolor[rgb]{0.77, 0.87, 0.71}\textbf{236.68} & 6 \\ \hline
        s13207 & 1988.25 & 332.29 & 1 & 2002.24 & \cellcolor[rgb]{0.77, 0.87, 0.71}\textbf{310.50} & 8 & 2023.24 & \cellcolor[rgb]{0.77, 0.87, 0.71}\textbf{326.79} & 7 & 2587.78 & 2070.82 & 1 & \cellcolor[rgb]{0.77, 0.87, 0.71}\textbf{2423.08} & \cellcolor[rgb]{0.77, 0.87, 0.71}\textbf{1287.35} & 8 & \cellcolor[rgb]{0.77, 0.87, 0.71}\textbf{2574.94} & \cellcolor[rgb]{0.77, 0.87, 0.71}\textbf{1457.41} & 7 \\ \hline
        s15850 & 2321.84 & 769.57 & 2 & \cellcolor[rgb]{0.77, 0.87, 0.71}\textbf{2281.71} & 794.03 & 8 & \cellcolor[rgb]{0.77, 0.87, 0.71}\textbf{2284.28} & 812.92 & 7 & 3021.21 & 705.81 & 1 & \cellcolor[rgb]{0.77, 0.87, 0.71}\textbf{2750.84} & \cellcolor[rgb]{0.77, 0.87, 0.71}\textbf{687.76} & 8 & \cellcolor[rgb]{0.77, 0.87, 0.71}\textbf{2751.07} & \cellcolor[rgb]{0.77, 0.87, 0.71}\textbf{681.86} & 4 \\ \hline
        s38584 & 7463.56 & 426.97 & 3 & 7504.85 & \cellcolor[rgb]{0.77, 0.87, 0.71}\textbf{421.81} & 8 & 7512.55 & \cellcolor[rgb]{0.77, 0.87, 0.71}\textbf{420.06} & 6 & 7882.53 & 462.28 & 2 & 7888.36 & \cellcolor[rgb]{0.77, 0.87, 0.71}\textbf{423.59} & 8 & \cellcolor[rgb]{0.77, 0.87, 0.71}\textbf{7851.97} & \cellcolor[rgb]{0.77, 0.87, 0.71}\textbf{424.06} & 6 \\ \hline
        s5378 & 814.61 & 250.80 & 1 & \cellcolor[rgb]{0.77, 0.87, 0.71}\textbf{804.12} & 260.83 & 4 & \cellcolor[rgb]{0.77, 0.87, 0.71}\textbf{811.35} & 259.13 & 3 & 1004.50 & 192.86 & 1 & \cellcolor[rgb]{0.77, 0.87, 0.71}\textbf{899.76} & 255.16 & 4 & \cellcolor[rgb]{0.77, 0.87, 0.71}\textbf{899.29} & 255.16 & 4 \\ \hline
        s9234 & 1108.08 & 330.84 & 1 & \cellcolor[rgb]{0.77, 0.87, 0.71}\textbf{1081.72} & 340.53 & 6 & 1111.35 &\cellcolor[rgb]{0.77, 0.87, 0.71}\textbf{ 283.11} & 6 & 1459.87 & 261.79 & 1 & 1472.93 & \cellcolor[rgb]{0.77, 0.87, 0.71}\textbf{218.84} & 4 & 1463.83 & \cellcolor[rgb]{0.77, 0.87, 0.71}\textbf{219.01} & 2 \\ \hline
        sin & 4377.27 & 2610.08 & 4 & 4446.08 & 2625.40 & 9 & 4393.60 & 2618.41 & 8 & 10839.12 & 3932.97 & 5 & \cellcolor[rgb]{0.77, 0.87, 0.71}\textbf{9479.8} & \cellcolor[rgb]{0.77, 0.87, 0.71}\textbf{3433.33} & 5 & \cellcolor[rgb]{0.77, 0.87, 0.71}\textbf{9694.88} & \cellcolor[rgb]{0.77, 0.87, 0.71}\textbf{3519.91} & 5 \\ \hline
        sqrt & 16676.49 & 78936.76 & 9 & \cellcolor[rgb]{0.77, 0.87, 0.71}\textbf{16547.72} & \cellcolor[rgb]{0.77, 0.87, 0.71}\textbf{77219.29} & 11 & 16759.54 & \cellcolor[rgb]{0.77, 0.87, 0.71}\textbf{77502.61} & 20 & 46959.03 & 84295.30 & 25 & 46880.18 & \cellcolor[rgb]{0.77, 0.87, 0.71}\textbf{82075.75} & \cellcolor[rgb]{0.77, 0.87, 0.71}\textbf{24} & 47168.05 & \cellcolor[rgb]{0.77, 0.87, 0.71}\textbf{82546.56} & 31 \\ \hline
        square & 15278.67 & 3474.66 & 5 & 15680.38 & \cellcolor[rgb]{0.77, 0.87, 0.71}\textbf{3389.66} & 19 & 15737.30 & \cellcolor[rgb]{0.77, 0.87, 0.71}\textbf{3350.52} & 18 & 22591.77 & 801.41 & 7 & \cellcolor[rgb]{0.77, 0.87, 0.71}\textbf{21474.36} &\cellcolor[rgb]{0.77, 0.87, 0.71}\textbf{ 745.97} & 7 & \cellcolor[rgb]{0.77, 0.87, 0.71}\textbf{21502.12} & \cellcolor[rgb]{0.77, 0.87, 0.71}\textbf{752.64} & \cellcolor[rgb]{0.77, 0.87, 0.71}\textbf{6} \\ \hline
        tv80 & 4843.36 & 776.71 & 4 & \cellcolor[rgb]{0.77, 0.87, 0.71}\textbf{4763.58} & 873.94 & 7 & \cellcolor[rgb]{0.77, 0.87, 0.71}\textbf{4788.54} & 866.11 & 7 & 5735.66 & 761.48 & 4 & \cellcolor[rgb]{0.77, 0.87, 0.71}\textbf{5632.31} & \cellcolor[rgb]{0.77, 0.87, 0.71}\textbf{709.92} & 7 & \cellcolor[rgb]{0.77, 0.87, 0.71}\textbf{5621.35} & \cellcolor[rgb]{0.77, 0.87, 0.71}\textbf{709.58} & 6 \\ \hline
        vga\_lcd & 58923.03 & 3173.61 & 71 & 60094.09 & \cellcolor[rgb]{0.77, 0.87, 0.71}\textbf{3149.33} & \cellcolor[rgb]{0.77, 0.87, 0.71}\textbf{16} & 60091.06 & \cellcolor[rgb]{0.77, 0.87, 0.71}\textbf{3149.33} & \cellcolor[rgb]{0.77, 0.87, 0.71}\textbf{18} & 93101.12 & 5919.01 & 44 & \cellcolor[rgb]{0.77, 0.87, 0.71}\textbf{83900.55} & \cellcolor[rgb]{0.77, 0.87, 0.71}\textbf{5509.76} & \cellcolor[rgb]{0.77, 0.87, 0.71}\textbf{23} & \cellcolor[rgb]{0.77, 0.87, 0.71}\textbf{83893.79} & \cellcolor[rgb]{0.77, 0.87, 0.71}\textbf{5509.76} & \cellcolor[rgb]{0.77, 0.87, 0.71}\textbf{18} \\ \hline
        voter & 8481.13 & 878.13 & 2 & \cellcolor[rgb]{0.77, 0.87, 0.71}\textbf{8427.24} & 928.81 & 14 & 8560.21 & 889.26 & 22 & 17275.55 & 966.30 & 7 & \cellcolor[rgb]{0.77, 0.87, 0.71}\textbf{13944.08} & \cellcolor[rgb]{0.77, 0.87, 0.71}\textbf{871.07} & 37 & \cellcolor[rgb]{0.77, 0.87, 0.71}\textbf{14355.12} & \cellcolor[rgb]{0.77, 0.87, 0.71}\textbf{878.20} & 84 \\ \hline
        core & 13751.39 & 336.07 & 11 & 13952.24 & 344.20 & \cellcolor[rgb]{0.77, 0.87, 0.71}\textbf{8} & 13979.30 & 343.19 & 47 & 17938.53 & 303.25 & 7 & \cellcolor[rgb]{0.77, 0.87, 0.71}\textbf{17799.03} & \cellcolor[rgb]{0.77, 0.87, 0.71}\textbf{292.46} & 9 & \cellcolor[rgb]{0.77, 0.87, 0.71}\textbf{17699.65} & \cellcolor[rgb]{0.77, 0.87, 0.71}\textbf{294.34} & 8 \\ \hline
        aes\_area & 2999.05 & 820.15 & 3 & 3015.61 & \cellcolor[rgb]{0.77, 0.87, 0.71}\textbf{652.68} & 8 & 3056.67 & \cellcolor[rgb]{0.77, 0.87, 0.71}\textbf{664.70} & 7 & 4915.91 & 743.96 & 3 & \cellcolor[rgb]{0.77, 0.87, 0.71}\textbf{4910.08} &\cellcolor[rgb]{0.77, 0.87, 0.71}\textbf{ 743.57} & 8 & \cellcolor[rgb]{0.77, 0.87, 0.71}\textbf{4884.18} & 776.98 & 7 \\ \hline
        pci\_ctrl & 513.92 & 294.54 & 1 & 514.62 & \cellcolor[rgb]{0.77, 0.87, 0.71}\textbf{292.23} & 6 & 521.61 & \cellcolor[rgb]{0.77, 0.87, 0.71}\textbf{279.83} & 5 & 743.46 & 314.71 & 1 & \cellcolor[rgb]{0.77, 0.87, 0.71}\textbf{743.00} & \cellcolor[rgb]{0.77, 0.87, 0.71}\textbf{283.14} & 6 & \cellcolor[rgb]{0.77, 0.87, 0.71}\textbf{722.70} & \cellcolor[rgb]{0.77, 0.87, 0.71}\textbf{270.13} & 4 \\ \hline
        systemcaes & 6349.88 & 1045.12 & 3 & \cellcolor[rgb]{0.77, 0.87, 0.71}\textbf{6272.20} & \cellcolor[rgb]{0.77, 0.87, 0.71}\textbf{862.01} & 7 & \cellcolor[rgb]{0.77, 0.87, 0.71}\textbf{6267.30} & \cellcolor[rgb]{0.77, 0.87, 0.71}\textbf{873.25} & 5 & 7634.09 & 1216.68 & 4 & \cellcolor[rgb]{0.77, 0.87, 0.71}\textbf{7530.28} & \cellcolor[rgb]{0.77, 0.87, 0.71}\textbf{1121.65} & 5 & \cellcolor[rgb]{0.77, 0.87, 0.71}\textbf{7503.22} & \cellcolor[rgb]{0.77, 0.87, 0.71}\textbf{1079.17} & 5 \\ \hline
        Geomean & 4424.82 & 1482.92 & 3.78 & \cellcolor[rgb]{0.77, 0.87, 0.71}\textbf{4420.20} & \cellcolor[rgb]{0.77, 0.87, 0.71}\textbf{1442.35} & 9.01 & 4447.13 & \cellcolor[rgb]{0.77, 0.87, 0.71}\textbf{1463.33} & 9.76 & 7053.97 & 1404.68 & 4.46 & \cellcolor[rgb]{0.77, 0.87, 0.71}\textbf{6782.61} & \cellcolor[rgb]{0.77, 0.87, 0.71}\textbf{1287.35} & 9.48 & \cellcolor[rgb]{0.77, 0.87, 0.71}\textbf{6825.51} & \cellcolor[rgb]{0.77, 0.87, 0.71}\textbf{1336.16} & 9.08 \\ \hline
        Impr. &  &  &  & \cellcolor[rgb]{0.77, 0.87, 0.71}\textbf{0.11\%} & \cellcolor[rgb]{0.77, 0.87, 0.71}\textbf{2.74\%} &  & -0.44\% & \cellcolor[rgb]{0.77, 0.87, 0.71}\textbf{1.39\%} & ~ &  &  & ~ &\cellcolor[rgb]{0.77, 0.87, 0.71}\textbf{3.85\%}&\cellcolor[rgb]{0.77, 0.87, 0.71}\textbf{8.35\%} &~ & \cellcolor[rgb]{0.77, 0.87, 0.71}\textbf{3.24\%} &\cellcolor[rgb]{0.77, 0.87, 0.71}\textbf{4.88\%} &~\\ \hline
\end{tabular}}
\label{tab:all-results}
\vspace{-10pt}
\end{table*}

\question{\textbf{How effective is \deftitle for technology mapping?}}
In this work, we compare the effectiveness of our choice-based optimization with the \texttt{dch} operator in \texttt{ABC}. We categorize the experiments into two optimization objectives: delay-oriented and area-oriented.
For the delay-oriented optimization, we first apply sop balancing~\cite{mishchenko2011delay} optimizations using the sequence:
\texttt{if -g -K 6 -C 8; if -g -K 6 -C 8},
which is widely used for delay optimization in tools like OpenROAD~\cite{openroad} and Yosys~\cite{yosys}.
For the area-oriented optimization, we use the sequence:
\texttt{compress2rs; compress2rs},
which has been demonstrated to be a strong optimization approach for reducing area. Additionally, in both delay- and area-oriented experiments, we apply a single \texttt{compress2} to ensure fairness. This is because the \texttt{dch} operator internally uses both \texttt{{compress2}} and \texttt{compress} to construct snapshot intermediate networks, leveraging the results optimized by \texttt{compress2} as the main subject graph.
To ensure a fair comparison, we validate that the choice networks constructed from the subject graphs produced by our method and by \texttt{dch}, after removing all choices and directly performing mapping, yield identical mapping results. This eliminates any potential bias caused by differences in independent AIG optimizations and isolates the impact of choices on technology mapping.
For technology mapping, we use the following commands:
For delay-oriented mapping: \texttt{map;topo;upsize;dnsize;stime;}
For area-oriented mapping: replacing \texttt{map} with \texttt{map -a}.
The result is shown in~\Cref{tab:all-results}, where it can be observed that for the delay-oriented optimization, \deftitle outperforms \texttt{dch} by achieving a delay reduction on almost all designs, with a geomean improvement of 8.35\%. Moreover, since our method provides choice nodes specifically for \textit{MFFC}, it also leads to a geomean area recovery improvement of approximately 3.4\%. For the area-oriented optimization, \deftitle demonstrates that it achieves no degradation in area while still achieving a geomean delay reduction of 2.74\%. The reason our method is particularly advantageous in delay‑oriented optimization is that it is easier to identify larger \textit{MFFCs} as \textit{RCs}, making the structural changes of \textit{CCs} have a greater impact on mapping. Conversely, under area-oriented optimization, independent area-oriented techniques tend to shrink the size of \textit{MFFCs}, which reduces the impact of structural variants on optimization. As a result, our method is particularly well-suited for delay-oriented optimization, where larger \textit{MFFCs} and greater structural diversity amplify its benefits.

\question{\textbf{Does Representative Cones Selection Improve QoR? Do More Choices Always Lead to Better Results?}}
\begin{figure*}[!h]
    \centering
    \begin{subfigure}{0.48\textwidth}
        \centering
        \includegraphics[width=\textwidth]{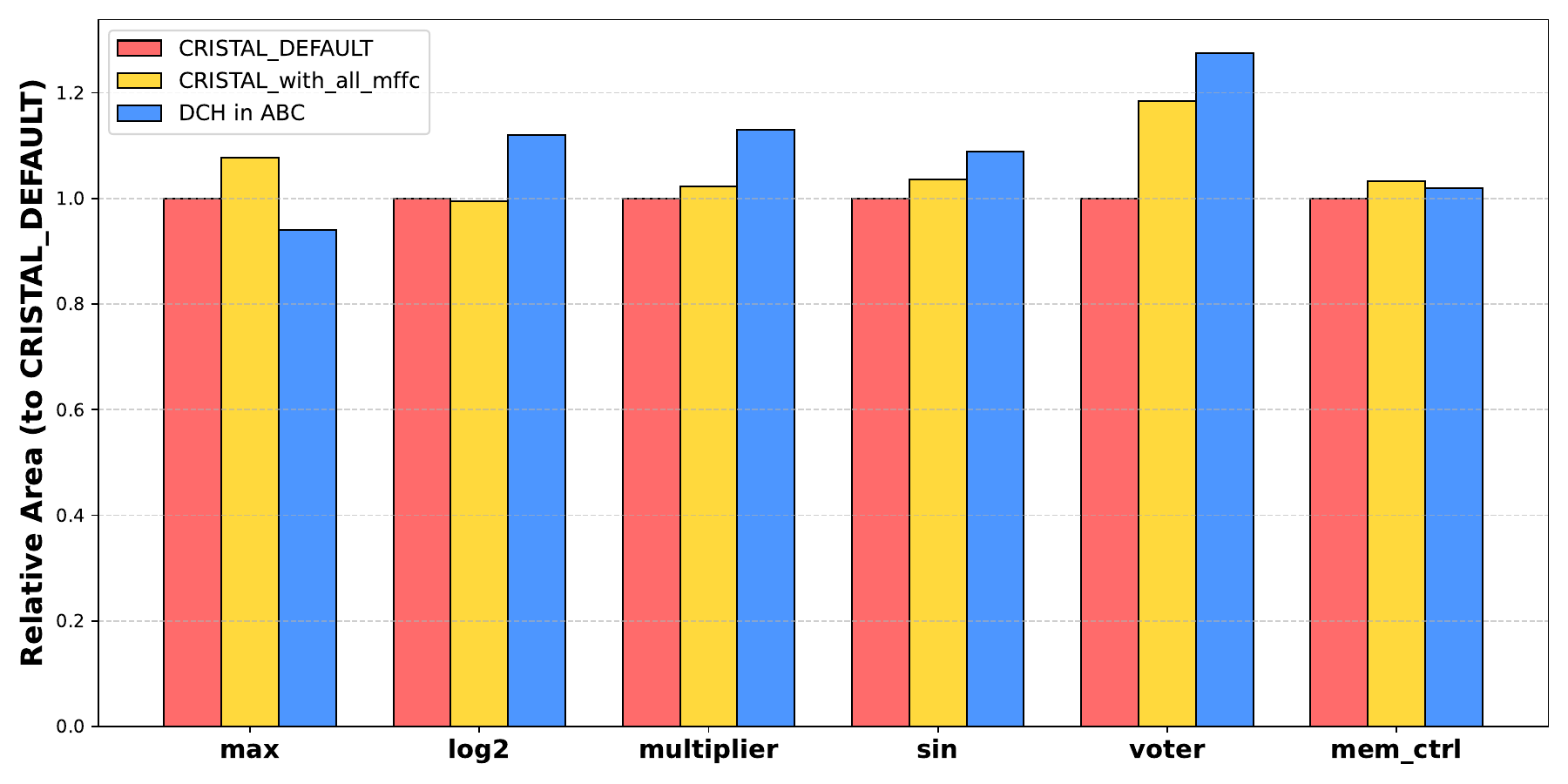}
        \caption{Normalized Area comparison.}
        \label{fig:area_bar}
    \end{subfigure}
    \hfill
    \begin{subfigure}{0.48\textwidth}
        \centering
        \includegraphics[width=\textwidth]{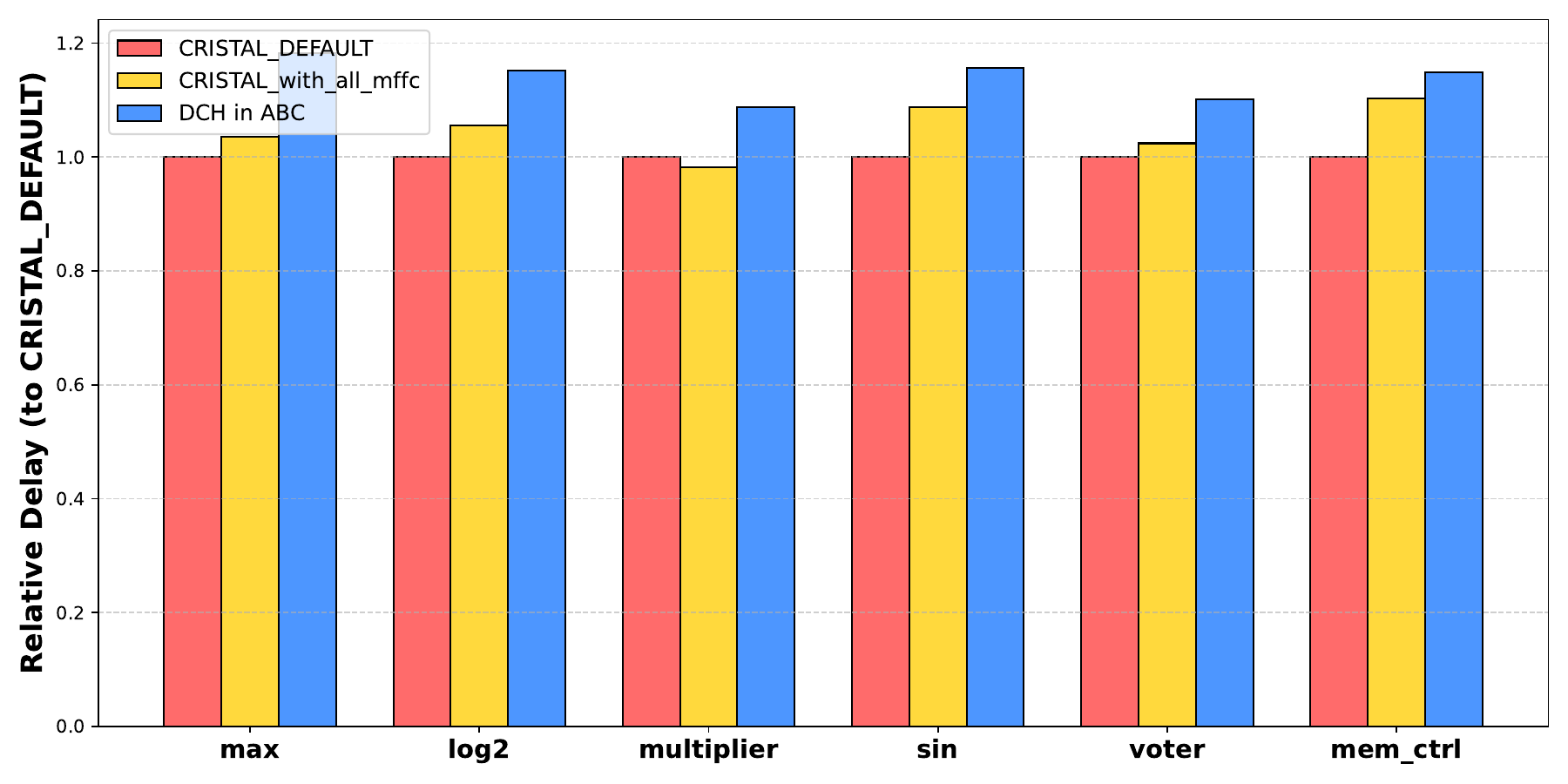}
        \caption{Normalized Delay comparison.}
        \label{fig:delay_bar}
    \end{subfigure}
    \caption{Normalized Delay and Area comparisons across different Representative Cones Selection methods.}
    \label{fig:exp2_normalized_bar}
\end{figure*}

We conducted comprehensive comparisons among three consfigurations: \deftitle with Representative Cones selection, \texttt{dch} in \texttt{ABC}, and \deftitle with exhaustive enumeration of all \textit{MFFCs}. For this analysis, we selected six medium-to-large EPFL benchmark circuits, compatibility with the previous delay-oriented optimization setup. As illustrated in~\Cref{fig:exp2_normalized_bar}, \deftitle with Representative Cones selection demonstrates marginally better results (4.38\% improvement in delay, 4.21\% in area) compared to the all-\textit{MFFC} baseline, while significantly outperforming the \texttt{dch} approach with 10.03\% improvement in delay and 9.99\% in area. These results empirically validate that strategic Representative Cones selection focusing on large \textit{MFFCs} and low-fanout cones effectively preserves critical optimization potential.

\begin{figure}[!h]
    \centering
    \includegraphics[width=0.48\textwidth]{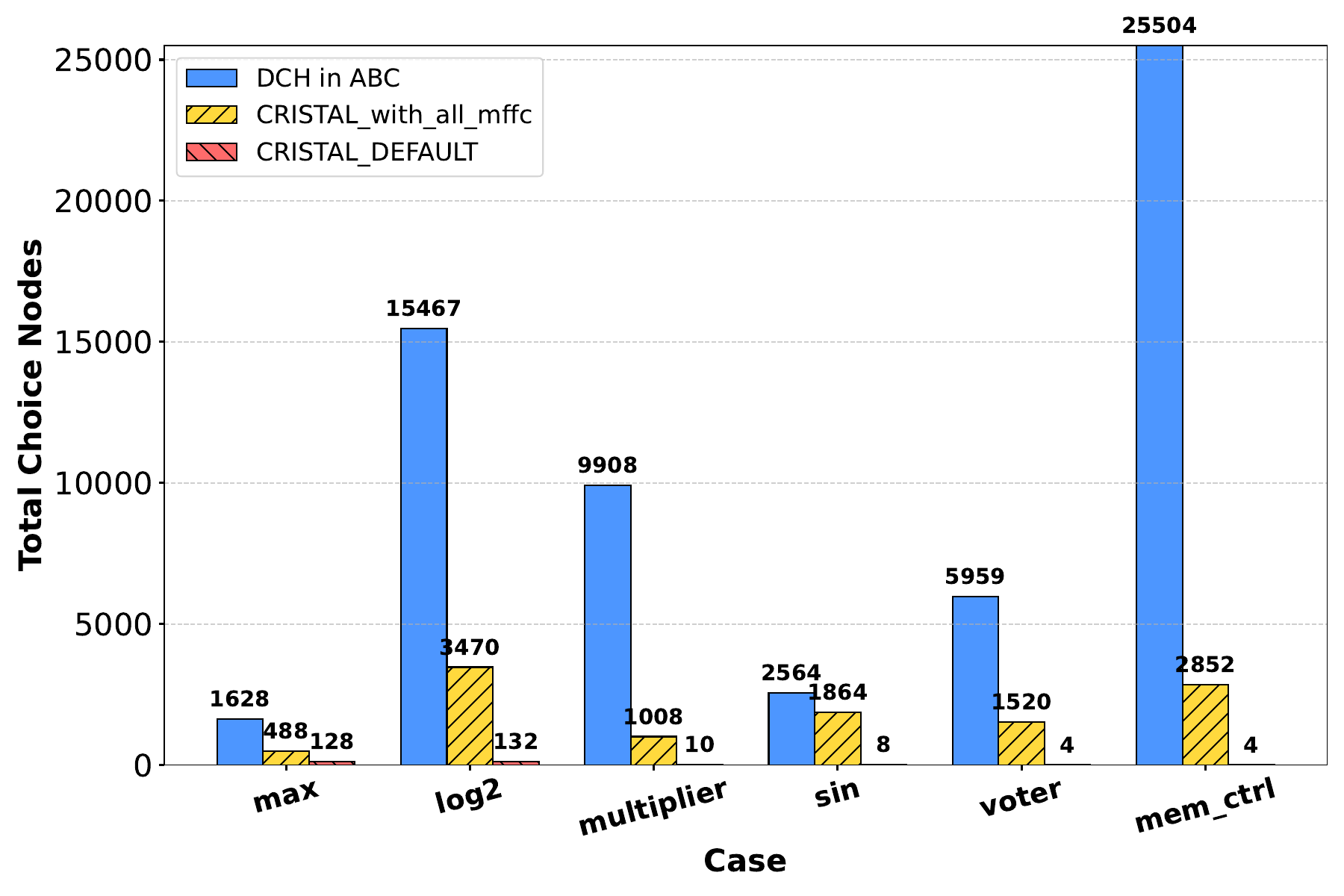}
    \caption{Total number of choices across different Representative Cones Selection methods.}
    \label{fig:choice_count}
\end{figure}

Furthermore, ~\Cref{fig:choice_count} highlights a striking contrast in choice space exploration: The \texttt{dch} generates 802× as many choice nodes as \deftitle. This significant disparity demonstrates that simply increasing the quantity of choices does not inherently lead to better quality-of-results (QoR). Instead, it is the selection of \textit{RCs} and the construction of structurally diverse Choice Cones, coupled with priority-based selection, that effectively identifies the most impactful optimization opportunities. These factors ensure that optimization focuses on high-value cones, avoiding unnecessary computational overhead while achieving superior results.

\question{\textbf{Can Priority-Ranked Filtering Help Mitigate Global Structural Bias?}}

To evaluate the efficacy of priority-ranked filtering in addressing global structural bias, we conduct an ablation study comparing technology mapping QoR with and without our hybrid priority-ranked scoring mechanism while maintaining identical equivalence classes and \textit{RCs}. As shown in~\Cref{tab:all-results}, the priority-ranked strategy yields a 1.43\% delay reduction with 0.60\% area degradation under area-oriented mapping, and achieves 3.47\% delay optimization with 0.61\% area improvement in delay-oriented mode. Notably, the method also reduces runtime by 5.95\% across benchmarks compared to the unfiltered approach, as fewer retained \textit{CCs} directly reduce the number of computations required during cut enumeration in Boolean matching.
Retaining all structural variants in the choice network overwhelms the covering phase with excessive cut patterns, which impacts the selection of heuristics such as area-flow cost. This inaccuracy is caused by the fanout count introduced by the choice network, leading to imprecise estimations. By contrast, priority-ranked filtering preserves only the top-3 structurally diverse and implementation-efficient \textit{CCs} per \textit{RC}, which constrains the search space to high-quality candidates. This selective pruning mitigates the "noise" from redundant variants. The mechanism’s success lies in its ability to balance structural diversity with implementation quality: structural exploration via equality saturation ensures sufficient coverage of the design space, while hybrid scoring ensures that only variants with strong local efficiency and global divergence advance to technology mapping. This dual strategy breaks the correlation between early greedy decisions and downstream structural constraints, effectively mitigating cascading suboptimality in the covering process.

\question{\textbf{Can \deftitle~Scale to Industrial-Scale Designs While Preserving Runtime Efficiency?}}

\begin{figure}[!h]
    \centering
    \includegraphics[width=0.5\textwidth]{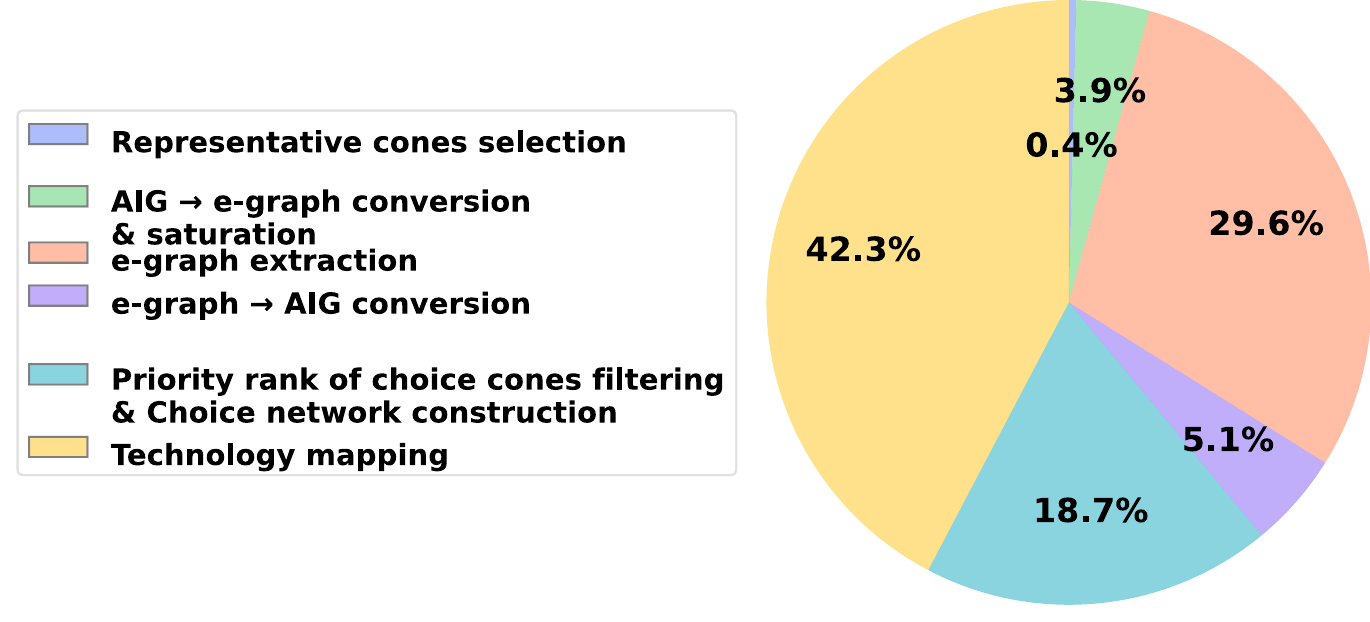}
    \caption{Runtime breakdown. }
\label{runtime}
\end{figure}

The overall runtime comparison is shown in~\Cref{tab:all-results}, where our detailed evaluation against \texttt{ABC}’s \texttt{dch} on the five largest benchmarks  demonstrates that \deftitle achieves 63.77\% faster execution, owing primarily to two  advantages: first, by eliminating verification overhead—\texttt{dch} enumerates a large number of candidate choices through extensive simulation, each of which must undergo SAT‑based equivalence verification, whereas \deftitle directly constructs provably equivalent \textit{CCs} via e‑graph rewriting and traditional optimization, obviating any SAT calls; and second, by reducing mapping complexity—\deftitle supplies fewer but higher‑quality choices (top‑3 \textit{CCs} per \textit{RC}).  For smaller designs, \deftitle incurs a modest overhead from data conversion and data transmission. As shown in~\Cref{runtime}, the runtime breakdown reveals: Representative Cone Selection (0.4\%), \textit{CCs} Generation (38.6\%), Priority‑Ranked Filtering (18.7\%), and Technology Mapping (42.3\%). Technology Mapping consumes the greatest share of time—nearly matching the combined runtime of all other stages involved in constructing the choice network. These performance benefits are driven by three key innovations: (1) an end‑to‑end Rust‑based, zero‑copy AIG‑to‑e‑graph parser that eliminates the intermediate conversions to equation or BLIF formats used in previous work~\cite{chen2024syn}~\cite{yin2025boole}; (2) leveraging 64‑core multithreading across all e‑graph operations for \textit{CCs} construction and (3) a lightweight, priority‑ranked \textit{CCs} filtering algorithm that delivers near‑linear scaling. Together, these optimizations make \deftitle both scalable and practical for industrial‑scale circuit synthesis, delivering predictable runtimes.



\section{Conclusion}
\label{sec:conclusion}
This paper presents \deftitle, a novel framework for constructing high-quality Boolean choice networks to mitigate structural bias in technology mapping. Unlike prior methods that passively aggregate functionally equivalent nodes from independent optimizations, \deftitle actively rethinks choice generation through three key innovations: (1) targeted selection of \emph{Representative Cones (RCs)} that significantly impact mapping quality; (2) a hybrid strategy that combines equality saturation-based structural mutation with traditional optimization-based mutation to generate diverse \emph{Choice Cones (CCs)}, breaking the homogeneity of local rewrites; and (3) a lightweight hybrid evaluation metric---integrating simulation-based fingerprinting, AND-gate disparity, and depth-fanout Pearson correlation---which enables fast structural dissimilarity assessment and the integration of independent and local mapping QoR evaluations to prioritize high-quality \emph{CCs}. Notably, \deftitle scales efficiently on industrial-scale circuits, attributed to parallel e-graph processing and selective choice pruning. By advocating "fewer but better" choices, \deftitle provides a scalable solution to choice network construction.



\section{acknowledgement}{This work is sponsored by the National Science Foundation under Grant No. CCF2403134, CCF2349670, CNS2349461, and CNS2229562.}



\newpage
\bibliography{refs}

\begin{thebibliography}{10}

\bibitem{amaru2015epfl}
L.~Amar{\'u}, P.-E. Gaillardon, and G.~De~Micheli.
\newblock The {EPFL} combinational benchmark suite.
\newblock In {\em IWLS}, number CONF, 2015.

\bibitem{MIG}
L.~Amarú, P.-E. Gaillardon, and G.~De~Micheli.
\newblock Majority-inverter graph: A new paradigm for logic optimization.
\newblock {\em TCAD}, 35(5):806--819, 2016.

\bibitem{Brglez1989}
F.~Brglez, D.~Bryan, and K.~Kozminski.
\newblock Combinational profiles of sequential benchmark circuits.
\newblock In {\em IEEE International Symposium on Circuits and Systems (ISCAS)}, pages 1929--1934. IEEE, 1989.

\bibitem{cai2025smoothe}
Y.~Cai, K.~Yang, C.~Deng, C.~Yu, and Z.~Zhang.
\newblock Smoothe: Differentiable e-graph extraction.
\newblock In {\em Proceedings of the 30th ACM International Conference on Architectural Support for Programming Languages and Operating Systems, Volume 1}, pages 1020--1034, 2025.

\bibitem{chatterjee2007algorithms}
S.~Chatterjee.
\newblock {\em On algorithms for technology mapping}.
\newblock University of California, Berkeley, 2007.

\bibitem{chatterjee2006reducing}
S.~Chatterjee, A.~Mishchenko, R.~K. Brayton, X.~Wang, and T.~Kam.
\newblock Reducing structural bias in technology mapping.
\newblock {\em IEEE Transactions on Computer-Aided Design of Integrated Circuits and Systems}, 25(12):2894--2903, 2006.

\bibitem{chen2025emorphicscalableequalitysaturation}
C.~Chen, G.~HU, C.~Yu, Y.~Ma, and H.~Zhang.
\newblock E-morphic: Scalable equality saturation for structural exploration in logic~synthesis, 2025.

\bibitem{chen2024syn}
C.~Chen, G.~Hu, D.~Zuo, C.~Yu, Y.~Ma, and H.~Zhang.
\newblock E-syn: E-graph rewriting with technology-aware cost functions for logic synthesis.
\newblock In {\em Proceedings of the 61st ACM/IEEE Design Automation Conference}, pages 1--6, 2024.

\bibitem{cong}
G.~Chen and J.~Cong.
\newblock Simultaneous logic decomposition with technology mapping in fpga designs.
\newblock In {\em Proceedings of the 2001 ACM/SIGDA ninth international symposium on Field programmable gate arrays}, pages 48--55, 2001.

\bibitem{cheng2024seer}
J.~Cheng, S.~Coward, L.~Chelini, R.~Barbalho, and T.~Drane.
\newblock Seer: Super-optimization explorer for high-level synthesis using e-graph rewriting.
\newblock In {\em Proceedings of the 29th ACM International Conference on Architectural Support for Programming Languages and Operating Systems, Volume 2}, pages 1029--1044, 2024.

\bibitem{openroad}
V.~A. Chhabria, W.~Jiang, A.~B. Kahng, R.~Liang, H.~Ren, S.~S. Sapatnekar, and B.-Y. Wu.
\newblock Openroad and circuitops: Infrastructure for ml eda research and education.
\newblock In {\em 2024 IEEE 42nd VLSI Test Symposium (VTS)}, pages 1--4, 2024.

\bibitem{clark2016asap7}
L.~T. Clark, V.~Vashishtha, L.~Shifren, A.~Gujja, S.~Sinha, B.~Cline, C.~Ramamurthy, and G.~Yeric.
\newblock Asap7: A 7-nm finfet predictive process design kit.
\newblock {\em Microelectronics Journal}, 53:105--115, 2016.

\bibitem{cohen2009pearson}
I.~Cohen, Y.~Huang, J.~Chen, J.~Benesty, J.~Benesty, J.~Chen, Y.~Huang, and I.~Cohen.
\newblock Pearson correlation coefficient.
\newblock {\em Noise reduction in speech processing}, pages 1--4, 2009.

\bibitem{cong1999cut}
J.~Cong, C.~Wu, and Y.~Ding.
\newblock Cut ranking and pruning: Enabling a general and efficient fpga mapping solution.
\newblock In {\em Proceedings of the 1999 ACM/SIGDA seventh international symposium on Field programmable gate arrays}, pages 29--35, 1999.

\bibitem{coward2022automatic}
S.~Coward, G.~A. Constantinides, and T.~Drane.
\newblock Automatic datapath optimization using e-graphs.
\newblock In {\em 2022 IEEE 29th Symposium on Computer Arithmetic (ARITH)}, pages 43--50. IEEE, 2022.

\bibitem{coward2023automating}
S.~Coward, G.~A. Constantinides, and T.~Drane.
\newblock Automating constraint-aware datapath optimization using e-graphs.
\newblock In {\em DAC}, 2023.

\bibitem{coward2023datapath}
S.~Coward, E.~Morini, B.~Tan, T.~Drane, and G.~A. Constantinides.
\newblock Datapath verification via word-level e-graph rewriting.
\newblock In {\em 2023 Formal Methods in Computer-Aided Design (FMCAD)}, pages 92--100. IEEE, 2023.

\bibitem{gao2010survey}
X.~Gao, B.~Xiao, D.~Tao, and X.~Li.
\newblock A survey of graph edit distance.
\newblock {\em Pattern Analysis and applications}, 13:113--129, 2010.

\bibitem{sat}
E.~I. Goldberg, M.~R. Prasad, and R.~K. Brayton.
\newblock Using sat for combinational equivalence checking.
\newblock In {\em Proceedings Design, Automation and Test in Europe. Conference and Exhibition 2001}, pages 114--121. IEEE, 2001.

\bibitem{grosnit2023lightweight}
A.~Grosnit, M.~Zimmer, R.~Tutunov, X.~Li, L.~Chen, F.~Yang, M.~Yuan, and H.~Bou-Ammar.
\newblock Lightweight structural choices operator for technology mapping.
\newblock In {\em 2023 60th ACM/IEEE Design Automation Conference (DAC)}, pages 1--6. IEEE, 2023.

\bibitem{XIG}
W.~Haaswijk, M.~Soeken, L.~Amarù, P.-E. Gaillardon, and G.~De~Micheli.
\newblock A novel basis for logic rewriting.
\newblock In {\em ASP-DAC}, pages 151--156, 2017.

\bibitem{hu2025mixedstructuralchoiceoperator}
Z.~Hu, H.~Pan, Y.~Xia, L.~Wang, and Z.~Chu.
\newblock Mixed structural choice operator: Enhancing technology mapping with heterogeneous representations, 2025.

\bibitem{lehman1997logic}
E.~Lehman, Y.~Watanabe, J.~Grodstein, and H.~Harkness.
\newblock Logic decomposition during technology mapping.
\newblock {\em IEEE Transactions on Computer-Aided Design of Integrated Circuits and Systems}, 16(8):813--834, 1997.

\bibitem{booleanmatching}
F.~Mailhot and G.~De~Micheli.
\newblock Technology mapping using boolean matching.
\newblock In {\em Proc. Eur. Conf. Design Automation}, pages 180--185, 1990.

\bibitem{history}
A.~Mishchenko and R.~Brayton.
\newblock Recording synthesis history for sequential verification.
\newblock In {\em 2008 Formal Methods in Computer-Aided Design}, pages 1--8. IEEE, 2008.

\bibitem{mishchenko2011delay}
A.~Mishchenko, R.~Brayton, S.~Jang, and V.~Kravets.
\newblock Delay optimization using sop balancing.
\newblock In {\em 2011 IEEE/ACM International Conference on Computer-Aided Design (ICCAD)}, pages 375--382. IEEE, 2011.

\bibitem{aig2006}
A.~Mishchenko, S.~Chatterjee, and R.~Brayton.
\newblock Dag-aware aig rewriting a fresh look at combinational logic synthesis.
\newblock In {\em Proceedings of the 43rd annual Design Automation Conference}, pages 532--535, 2006.

\bibitem{mishchenko2006improvements}
A.~Mishchenko, S.~Chatterjee, and R.~Brayton.
\newblock Improvements to technology mapping for lut-based fpgas.
\newblock In {\em Proceedings of the 2006 ACM/SIGDA 14th international symposium on Field programmable gate arrays}, pages 41--49, 2006.

\bibitem{ustun2022impress}
E.~Ustun, I.~San, J.~Yin, C.~Yu, and Z.~Zhang.
\newblock Impress: Large integer multiplication expression rewriting for fpga hls.
\newblock In {\em 2022 IEEE 30th Annual International Symposium on Field-Programmable Custom Computing Machines (FCCM)}, pages 1--10. IEEE, 2022.

\bibitem{egg}
M.~Willsey, C.~Nandi, Y.~R. Wang, O.~Flatt, Z.~Tatlock, and P.~Panchekha.
\newblock Egg: Fast and extensible equality saturation.
\newblock In {\em POPL}, volume~5, pages 1--29, New York, NY, USA, 2021. ACM.

\bibitem{yosys}
C.~Wolf.
\newblock Yosys open synthesis suite.
\newblock \url{https://yosyshq.net/yosys/}.

\bibitem{yin2025boole}
J.~Yin, Z.~Song, C.~Chen, Q.~Hu, and C.~Yu.
\newblock Boole: Exact symbolic reasoning via boolean equality saturation.
\newblock {\em arXiv preprint arXiv:2504.05577}, 2025.

\end{thebibliography}
\balance

\end{document}